\documentclass[a4paper,twocolumn,fleqn,nofootinbib]{revtex4-1}

\usepackage{amsmath}             
\usepackage{txfonts}               
 \usepackage{bm}                  
\usepackage{graphicx}

\usepackage[normalem]{ulem} 
\usepackage{units} 
\usepackage{float} 

\def\O{\mathcal{O}}

\def\V{\mathcal{V}}

\def\J{\mathcal{J}}

\def\N{\mathcal{N}}
\def\enr{\mathcal{E}}

\def\nablab{{\bm \nabla}}

\AtBeginDocument{\mathcode`v=\varv} 

\AtBeginDocument{\mathcode`v=\varv} 
\newcommand{\vertapprox}{\rotatebox{90}{$\approx$}}

\makeatletter
\def\p@subsection{}
\makeatother
\makeatletter
\AtBeginDocument{
\g@addto@macro{\appendix}{}
}
\makeatother

\begin{document}

\title{Representation and modeling of charged particle distributions in tokamaks}

\author{Andreas~Bierwage$^{1}$}
\email[\it Author's electronic address: ]{bierwage.andreas@qst.go.jp}
\author{Michael~Fitzgerald$^{2}$}
\author{Philipp~Lauber$^{3}$}
\author{Mirko~Salewski$^{4}$}
\author{Yevgen~Kazakov$^{5}$}
\author{\v{Z}iga~\v{S}tancar$^{6,2}$}

\affiliation{
  $^{1}$QST, Rokkasho and Naka Fusion Institutes, Japan \\
  $^{2}$\mbox{Culham Centre for Fusion Energy of UKAEA, Culham Science Centre, Abingdon, United Kingdom} \\
  $^{3}$Max-Planck-Institut f\"{u}r Plasmaphysik, Garching, Germany \\
  $^{4}$\mbox{Department of Physics, Technical University of Denmark, Kgs.\ Lyngby, Denmark} \\
  $^{5}$\mbox{Laboratory for Plasma Physics, LPP-ERM/KMS, Partner in the Trilateral Euregio Cluster (TEC), Brussels, Belgium} \\
  $^{6}$Jo\v{z}ef Stefan Institute, Ljubljana, Slovenia}

\begin{abstract}
Experimental diagnostics, analysis tools and simulations represent particle distributions in various forms and coordinates. Algorithms to manage these data are needed on platforms like the ITER Integrated Modelling \& Analysis Suite (IMAS), performing tasks such as archiving, modeling, conversion and visualization. A method that accomplishes some of the required tasks for distributions of charged particles with arbitrarily large magnetic drifts in axisymmetric tokamak geometry is described here. Given a magnetic configuration, we first construct a database of guiding center orbits, which serves as a basis for representing particle distributions. The orbit database contains the geometric information needed to perform conversions between arbitrary coordinates, modeling tasks, and resonance analyses. Using that database, an imported or newly modeled distribution is mapped to an exact equilibrium, where the dimensionality is reduced to three constants of motion (CoM). The orbit weight is uniquely given when the input is a true distribution: one that measures the number of physical particles per unit of phase space volume. Less ideal inputs, such as distributions estimated without drifts, or models of particle sources, can also be processed. As an application example, we reconstruct the drift-induced features of a distribution of fusion-born alpha particles in a large tokamak, given only a birth profile, which is not a function of the alpha's CoM. Repeated back-and-forth transformations between CoM space and energy-pitch-cylinder coordinates are performed for verification and as a proof of principle for IMAS.{\looseness=-1}
\end{abstract}



\maketitle  

\thispagestyle{empty}
\everypar{\looseness=-1} 

\makeatletter
\def\l@subsection#1#2{}
\def\l@subsubsection#1#2{}
\makeatother

\tableofcontents


\section{Introduction and workflow}
\label{sec:intro}

\subsection{Motivation, purpose and scope}

The ITER Integrated Modelling \& Analysis Suite (IMAS) \cite{PinchesFEC18} offers various ways to store particle distributions for the study of magnetically confined fusion plasmas. Different representations include different choices of coordinates and different discretization methods (mesh grids or marker particles). The Energetic Particle Topical Group of the International Tokamak Physics Activity (ITPA) is currently driving an action to equip IMAS with tools to model and convert distributions of fast particles between different representations that arise in experimental and computational work.

In experimental work, the observed position space of an energetic particle diagnostic is often given by the line-of-sight of the diagnostic in cylinder coordinates $(R,z,\zeta)$, where $R$ is the major radius, $z$ the height and $\zeta$ the toroidal angle. In tokamaks, $\zeta$ is often an ignorable symmetry coordinate. For many diagnostics, the observed velocity-space is given by weight functions that are derived from energy and momentum conservation \cite{Salewski15, Salewski16, Salewski18a, Salewski19}. The diagnostic weights are often expressed as functions of kinetic energy $E = Mv^2/2$ and velocity pitch $\lambda = v_\parallel/v$, where $v$ is the magnitude of the velocity vector ${\bm v}$, $v_\parallel$ its component parallel to the magnetic field vector ${\bm B}$, and $M$ the particle mass. Diagnosticians hence have a good understanding of the observation regions of diagnostics in phase space $(E,\lambda,R,z)$. Synthetic diagnostics calculating expected measurements based on numerical simulations are usually based on such representations of local velocity distributions.

Given the uncertainties of experimental measurements, no distinction is usually made between the gyro-averaged distribution of physical particles and the distribution of their guiding centers (GC), at least in the plasma core. The modeling of distribution functions is therefore often limited to GC statistics (except near the wall where the geometry of gyro-orbits matters for the interpretation of loss diagnostics and for the estimation of heat loads on plasma-facing hardware).

If one assumes that the plasma is not only toroidally symmetric but also quiescent, the distribution functions of confined charged particles and their GCs can be taken to depend only on three constants of motion (CoM). Analyses of resonant instabilities are usually performed with respect to such an equilibrated reference state, where the orbit time $\tau$, gyrophase $\xi$ and toroidal angle $\zeta$ are ignorable symmetry coordinates. The respective conserved actions --- namely, the kinetic energy $E$, magnetic moment $\mu$, and canonical toroidal angular momentum $P_\zeta$ --- combined with an index $\sigma$ specifying the sign of $v_\parallel$ on passing orbits, constitute a natural set of CoM coordinates $(E,\mu,P_\zeta;\sigma)$ that is common in theoretical analyses, although other useful (and equivalent) sets of CoM exist. Recently, an experimental measurement of a fast ion distribution function in three-dimensional (3-D) CoM space was demonstrated for fast-ion D-$\alpha$ spectroscopy \cite{Stagner22}. Efforts are underway to apply this orbit tomography method \cite{StagnerPhdThesis} to other fast-ion diagnostics \cite{Jaerleblad21}.

\begin{figure}
[tb]
\centering\vspace{-0.1cm}
\includegraphics[width=0.47\textwidth]{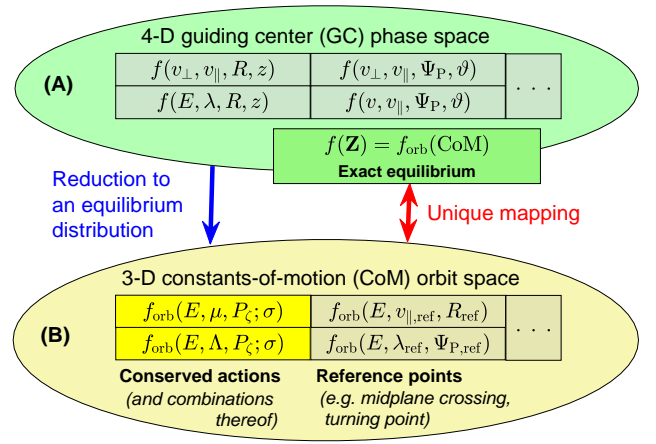}
\caption{Examples of coordinate representations and transformations (arrows) of a toroidally symmetric distribution function. We require that every 4-D distribution $f({\bm Z})$ be reduced to an exact equilibrium $f_{\rm orb}({\rm CoM})$, so that the mapping between 4-D GC phase space and 3-D CoM orbit space becomes unique and readily reversible. --- Position coordinates: major radius $R$; height $z$; poloidal magnetic flux $\Psi_{\rm P}(R,z)$; poloidal angle $\vartheta$. --- Velocity coordinates: GC velocity components $v_\perp$ and $v_\parallel$ relative to the magnetic field ${\bm B}$; kinetic energy $E = Mv^2/2$ with $v^2 = v_\parallel^2 + v_\perp^2$; pitch $\lambda = v_\parallel/v$; sign $\sigma = v_\parallel J_\parallel/|v_\parallel J_\parallel|$ relative to the parallel current density $J_\parallel$. --- Mixed coordinates: magnetic moment $\mu = Mv_\perp^2/(2B)$; conserved pitch $\Lambda = \mu B_0/E$; canonical toroidal angular momentum $P_\zeta$; coordinates $\Psi_{\rm P,ref}$, $\lambda_{\rm ref}$, etc., evaluated at a certain reference point, such as a midplane crossing or a turning point.}
\label{fig:transform}
\end{figure}

In order to make predictions of a measurement based on a stability calculation or carry out stability analyses based on measured data, we must be able to transform distribution functions between 3-D CoM space and 4-D representations given in various coordinates as illustrated in Fig.~\ref{fig:transform}. Related problems have been tackled to various degrees by different research groups with different codes and methods. For instance, {\tt TRANSP/NUBEAM} \cite{BreslauTRANSP17} and {\tt LIGKA/HAGIS} \cite{LauberFEC20} employ binning and smoothing techniques to map distributions represented by marker particles onto a mesh and compute their gradients in CoM space. While TRANSP/NUBEAM uses Monte-Carlo sampling, LIGKA/HAGIS uses a database of GC orbits for all particle species. Related activities were reported by the {\tt ASCOT} \cite{TholerusASCOT17} and {\tt MEGA} \cite{BierwageVSTART12} groups. These methods have been used for many years, but detailed documentations are rarely published. The above references \cite{BreslauTRANSP17,LauberFEC20,TholerusASCOT17,BierwageVSTART12} all point to presentations given at technical meetings. Useful elements can be found in many papers, but it is necessary to assemble that information into a complete goal-oriented workflow.

Thus motivated, this paper documents the workflow used in our computer code {\tt VisualStart} \cite{BierwageVSTART12, Bierwage12a}. Since 2009, we have used this code to initialize hybrid simulations with modeled or numerically computed beam ion distributions \cite{Bierwage13a, Bierwage14a, Bierwage17a, Bierwage18} and to analyze their resonant interactions with Alfv\'{e}n waves \cite{Bierwage14b, Bierwage16a, Bierwage16b}. {\tt VisualStart} represents GC distributions using marker particles, which can be directly used in full-$f$ simulations\footnote{For instance, see Fig.~3 in Ref.~\protect\cite{Bierwage13a}, and Fig.~4 in Ref.~\protect\cite{Bierwage14a} for fast ion tails, and Fig.~4 in Ref.~\protect\cite{Bierwage17a} for a birth distribution.}
and yield a quiet start \cite{Bierwage12a}. The representation in CoM space with an underlying GC orbit database will also allow us to directly embed the distribution functions in general phase-space transport theories \cite{Zonca15b, Chen16, Falessi19, Zonca21}, both formally and technically.

Here we propose the orbit-based representation and modeling technique as a solution for ITER IMAS and other platforms that need to store and process distributions of charged particles in magnetically confined plasmas. The subject of smoothing is discussed briefly in the following section, but the documentation and verification of smoothing algorithms is left for future work, as is the associated problem of initializing delta-$f$ simulations that require the evaluation of gradients. We note that a similar and in many ways complementary effort has been undertaken by S.~Benjamin {\it et al}.~\cite{BenjaminBachelorThesis}.

\subsection{Basic rules for practical situations}

The theoretical foundations for transforming distribution functions are relatively straightforward, but it is also clear that compromises and creativity are needed when tackling real-world data with numerical techniques, where accuracy has to be traded for computational performance and where data may have varying degrees of quality and completeness. For instance, the singularities and topological boundaries of CoM space \cite{Rome79} require attention when cutting a mesh to define phase space volume elements. Experimentally measured distributions and numerically computed distributions are often incomplete and noisy, so some amount of modeling is needed, including smoothing, interpolation and extrapolation.

In view of these challenges, some basic rules should be followed in order to maintain the physical and numerical integrity of the data, independently of the particular workflow used to process distribution functions:
\begin{enumerate}
\item[(i)]  Smoothing, interpolation and extrapolation are best applied at the preparation stage, where a new distribution function is first constructed or imported. Such potentially irreversible manipulations should be minimized in all subsequent operations.

\item[(ii)]  Pay attention to the fact that, in toroidal geometry, resolution and noise levels inevitably vary in space and in time, because the mirror force and magnetic drifts cause GCs to be distributed nonuniformly along their orbits. In particular:
\begin{itemize}
\item  Inherent topological discontinuities and singularities, as found in CoM space, should be preserved; e.g., through mesh accumulation and by smoothing only around, not across trapped-passing boundaries and loss cones.

\item  The optimal representation method (maximizing accuracy) may differ between archiving a distribution function and using it in a simulation. While archiving merely requires adequate resolution, simulations should also minimize signal-noise correlations \cite{Bierwage12a}.
\end{itemize}
These issues are most pronounced in but not limited to cases with large magnetic drifts.

\item[(iii)]  Ensure that the distribution to be transformed is an exact equilibrium in the given magnetic configuration; i.e., its shape must not vary in time when evolved using unperturbed equations of motion.
\end{enumerate}

The equilibrium constraint (iii) is crucial, so it deserves further explanation: Any distribution function $f$ can be mapped to CoM space, where the result $f_{\rm orb}$ measures the time-averaged particle densities on unperturbed GC orbits. This operation is indicated by a blue arrow in Fig.~\ref{fig:transform}. However, when the original $f$ is not an equilibrium, this mapping operation is time-dependent and it is impractical to store the information needed to invert it. The equilibrium constraint reduces the effective dimensionality to three CoM by determining the longitudinal distribution of GCs along their orbit, which is now time-independent and with it all coordinate transformations as well. The coordinate transformations $f\leftrightarrow f_{\rm orb}$ can then be readily performed in any direction, as indicated by a red double arrow in Fig.~\ref{fig:transform}.

Nonequilibrium distributions are very likely to be encountered in IMAS, where one can expect to receive particle data from a measurement for which the magnetohydrodynamic (MHD) equilibrium is not exactly known or from a simulation where waves or magnetic islands were present. Enforcing the equilibrium constraint means that the input data are transformed into a distribution that is consistent with a modeled magnetic configuration, which generally differs somewhat from the conditions where the input data originated from.

\begin{figure*}
[tb]
\centering
\includegraphics[width=0.88\textwidth]{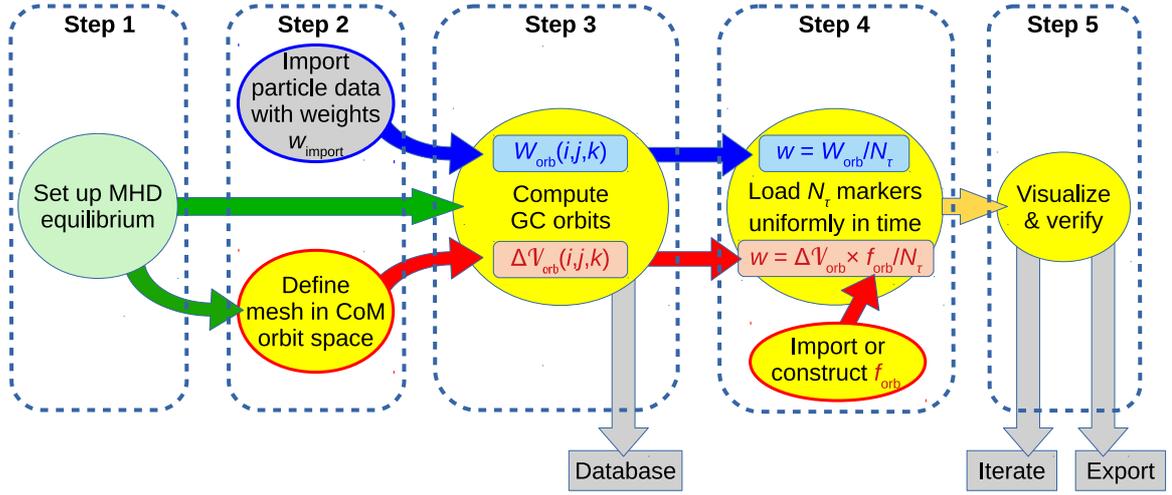}
\caption{Relevant part of the {\tt VisualStart} workflow in five steps as described in the text. This paper focuses on the tasks highlighted yellow.}
\label{fig:vstart}%
\end{figure*}

\subsection{Orbit-based representation and modeling}\vspace{-0.15cm}
\label{sec:intro_orb}

The above considerations suggest that it is useful and advantageous to represent distribution functions in CoM space with an underlying database of unperturbed GC orbits. With that geometric information, it is straightforward to reconstruct a 4-D distribution in arbitrary coordinates by loading a suitable number of markers uniformly in time along each GC orbit \cite{Bierwage12a}. Projections to various coordinates and views of user-defined sub-spaces with user-defined resolution can then be readily obtained by binning the markers on demand. This minimizes the amount of transformations and associated data corruption, as demanded in items (i) and (ii) above. Marker particles are a natural choice for representing fast particle distributions since the latter are often computed using particle codes. Moreover, by organizing the markers on orbit contours, one can identify and deal with inherent singularities and discontinuities as required in item (ii). Last but not least, by representing a distribution function in CoM orbit space, the equilibrium constraint (iii) is automatically enforced.

We note that the implementation and enforcement of the equilibrium constraint only requires a conversion to CoM coordinates: $f \rightarrow f_{\rm orb}$. The Jacobians required for such transformations can be determined using the Monte-Carlo method as implemented in {\tt TRANSP/NUBEAM} by Breslau \& Liu \cite{BreslauTRANSP17}. In this way, it is possible to avoid integration along GC orbits.

However, the computational cost associated with building an orbit database is more than repaid by the information one gains about the shape of the orbits, the longitudinal particle distribution, and the transit times. The transit times are needed for resonance analyses and their variation near trapped-passing boundaries can be used for mesh accumulation. The information about the orbit geometry is also useful for reduced models, for instance to capture finite-orbit-width effects \cite{Lauber18}, and it allows to define and make use of a variety of specialized GC coordinates, like the radii $R_{\rm m}$ where an orbit crosses the midplane. Last but not least, the information contained in the orbit database is also valuable for carrying out modeling tasks, such as the following example:

Given a model function $G_{\rm mdl}$ that may not be an exact solution of a kinetic equation for the GCs of physical particles, we wish to construct an equilibrium GC orbit distribution $f_{\rm orb}$. A common example are models of the form $G_{\rm mdl}(E,\lambda,\Psi_{\rm P})$ that contain measured radial profiles as functions of flux $\Psi_{\rm P}$ and a velocity distribution computed by a bounce-averaged Fokker-Planck code. In such a case, the orbit database allows to incorporate the effect of magnetic drifts {\it a posteriori} \cite{Bierwage17c}. There are, of course, various possibilities to perform this operation; in other words, there is some freedom to interpret the values of $G_{\rm mdl}$ as weights in an orbit distribution $f_{\rm orb}$.\footnote{Functions such as $G_{\rm mdl}({\bm Z})$ described in this example will here be called ``quasi-distributions'', because they do not measure the true number of objects of interest (here GCs) in a volume element ${\rm d}{\bm Z}$ but require more elaborate transformations than a mere conversion of coordinates. See Section~\protect\ref{sec:class}.}
An intuitive and physically meaningful choice is to perform an orbit-time average $\left<...\right>_{\rm orb}$ followed by a suitable normalization: $f_{\rm orb} = {\rm const.}\times\left<G_{\rm mdl}\right>_{\rm orb}$.

\subsection{Workflow}

The workflow described in this paper is shown schematically in Fig.~\ref{fig:vstart} and consists of five steps:
\begin{itemize}
\item  {\bf Step 1:} Define an axisymmetric magnetic field ${\bm B}$; e.g., using a dedicated MHD equilibrium code.

\item  {\bf Step 2:} Define samples in 3-D CoM orbit space, with cell indices $(i,j,k)$. The blue and red workflow branches in Fig.~\ref{fig:vstart} begin with different inputs:
\begin{enumerate}
\item[(blue)] Import marker particles from another code as in earlier works \protect\cite{Bierwage14a, Bierwage17a, Bierwage18}. The weights of the imported samples are reinterpreted as orbit weights: $w_{\rm import} \rightarrow W_{\rm orb}(i,j,k)$.

\item[(red)] Use a meshing algorithm to divide the CoM space into cells of size $\Delta\V_{\rm orb}(i,j,k)$. This path is described in detail in this paper.
\end{enumerate}

\item  {\bf Step 3:} Integrate along GC orbits starting from the samples defined in Step 2. Store in a database.

\item  {\bf Step 4:} Represent an equilibrium distribution function $f_{\rm orb}$ using weighted markers loaded on the orbits in the database. When loading $N_\tau$ markers uniformly in time, their weight is an $N_\tau$'s fraction of the orbit weight: $w = W_{\rm orb}/N_\tau$. In the case of the red workflow in Fig.~\ref{fig:vstart}, $W_{\rm orb} = \Delta\V_{\rm orb} \times f_{\rm orb}$ is the product of the volume element $\Delta\V_{\rm orb}$ and the desired phase space density function $f_{\rm orb}$. The latter may be imported or constructed from a model.

\item  {\bf Step 5:} Visualize and verify. Iterate if necessary; e.g., changing the resolution in CoM space (Step 2), the number of markers on the orbits, or the weight assignment (Step 4). Export when ready.
\end{itemize}

\noindent Note that MHD wave spectra can be obtained from the data available in Step 1, and resonance maps can be computed from the orbit database built in Step 3. Procedures for smoothing the distribution, computing phase space gradients and analyzing the stability of resonant modes remain to be added in or around Step 5.

\subsection{Outline}

Clear definitions of symbols and terminology are a prerequisite for successful conversion operations. Thus, we begin our treatise in Section~\ref{sec:class} with a discussion of different classes of distributions functions and how they are transformed. {\bf Step 1} of our workflow is part of Section~\ref{sec:geom}, where we describe the tokamak plasma that serves as a working example, define coordinates and normalizations. {\bf Step 2}, the construction of a phase space mesh, is covered in Section~\ref{sec:mesh}, where we also discuss orbit classes and the phase space topology in our scenario. The GC equations of motion we solve to construct the orbit database in {\bf Step 3} are given in Section~\ref{sec:eom}, and the formula for the orbit volume elements is derived in Section~\ref{sec:weight}. {\bf Steps 4 \& 5} are covered in Sections~\ref{sec:mdl} and \ref{sec:viz}, where we construct a distribution function $f$ from a model, visualize and analyze the result, and verify the integrity and accuracy of our scheme by transforming $f$ back and forth between two different representations. Modeling is a rich subject on its own, and our example is only meant to highlight the utility of an orbit database and to serve as a proof-of-principle for IMAS applications. The Appendices contain supplementary information on binning procedures and numerical accuracy.

\section{Classes of distributions and transformations}
\label{sec:class}

There are not only different coordinates and different numerical representations of distributions; the latter also come in different types whose identity must be clearly defined since it determines how they are transformed. Here, we distinguish between three types of distributions: (i) density functions, (ii) histograms, and (iii) quasi-distributions.

The first two can be characterized as follows:
\begin{enumerate}
\item[(i)] The symbols $f({\bm Z})$ represent {\it phase space densities}, which have units of $[{\rm m}^{-6}{\rm s}^{3}]$ and whose values are {\it independent} of the coordinates ${\bm Z}$ appearing in the argument, so $f({\bm Z}({\bm v},{\bm x})) = f({\bm v},{\bm x})$. The argument merely specifies the representation space. The same is true for velocity and volume integrals:
\begin{subequations}
\begin{align}
n({\bm x}) = \int{\rm d}^3{\bm v}\, f &: \text{density field}, \\
\nu({\bm v}) = \int_V{\rm d}^3{\bm x}\, f &: {\parbox{2.9cm}{velocity distribution \\ in a volume $V$.}}
\end{align}
\end{subequations}

\item[(ii)] The symbols $h({\bm Z})$ represent {\it histograms}, which have the units of the inverse volume element $|{\rm d}{\bm Z}|^{-1}$ of the coordinates used for ``binning'' (counting the objects of interest). Histograms are convenient to construct and integrate since no Jacobian is required (except for coordinate conversions). 
\end{enumerate}

\noindent The phase space density function $f$ is related to histogram functions $h$, $h'$, etc., as
\begin{equation}
f({\bm Z}) = \J^{-1}_{\bm Z} h({\bm Z}) = \J^{-1}_{{\bm Z}'} h'({\bm Z}') = ...
\label{eq:hist}
\end{equation}

\noindent where $\J_{\bm Z}$ is the Jacobian for the transformation from Cartesian coordinates ${\bm Z}_{\rm c} = ({\bm v},{\bm x}) = (v_x,v_y,v_z,x,y,z)$ to another arbitrary set of phase space coordinates ${\bm Z}$.

A phase space density function $f$ is an exact solution of a kinetic equation for the objects of interest; in our case, guiding centers (GC). In this work, we assume that all GC coordinates are defined such that they preserve the form of the kinetic equation
\begin{align}
0 =&\; (\partial_t + \dot{\bm Z}_{\rm c}\cdot\partial_{{\bm Z}_{\rm c}}) f
\\
=&\; \partial_t f + (\underbrace{\dot{\J}_{\rm gc}}\limits_0 {\bm Z}_{\rm gc} + \J_{\rm gc} \dot{\bm Z}_{\rm gc}) \cdot \frac{\partial_{{\bm Z}_{\rm gc}}f}{\J_{\rm gc}} \nonumber
\\
=&\; (\partial_t + \dot{\bm Z}_{\rm gc}\cdot\partial_{{\bm Z}_{\rm gc}})f.
\end{align}

\noindent This is the case when the Jacobian $\J_{\rm gc}$ for the transformation ${\bm Z}_{\rm c} \rightarrow {\bm Z}_{\rm gc}$ from Cartesian to GC coordinates satisfies the phase space conservation law
\begin{equation}
\dot{\J}_{\rm gc} \equiv \partial_t \J_{\rm gc} + \partial_{{\bm Z}_{\rm gc}}\cdot\left(\J_{\rm gc}\dot{\bm Z}_{\rm gc}\right) = 0.
\label{eq:djdt}
\end{equation}

Equation (\ref{eq:djdt}) is trivially satisfied for canonical coordinates, since their volume elements differ only by a constant factor, so $\J_{\rm gc} = {\rm const}$. All our CoM coordinates --- namely, $(E,\mu,P_\zeta;\sigma)$ and any equivalent set of GC orbit coordinates, such as $(v,v_{\parallel,{\rm ref}},R_{\rm ref})$ ---  are, by definition, conserved by the GC equations of motion that we solve. This means that, even when the equations of motion are not written in Hamiltonian form, our orbit coordinates combined with the triplet of ignorable angle variables $(\tau,\xi,\zeta)$  --- orbit time, gyrophase, toroidal angle --- can be considered to be canonical coordinates, whose Jacobians are constants satisfying Eq.~(\ref{eq:djdt}) down to the numerical accuracy of our particle-pushing algorithm.

A well-known set of noncanonical GC coordinates is $(\mu,v_\parallel,{\bm x}_{\rm gc})$, whose volume elements transform as
\begin{equation}
{\rm d}^3{\bm v}{\rm d}^3{\bm x} = \frac{2\pi B^*_\parallel}{M} {\rm d}\mu{\rm d}v_\parallel{\rm d}^3{\bm x}_{\rm gc}.
\label{eq:dvdx_noncan}
\end{equation}

\noindent The Jacobian factor $B^*_\parallel$ satisfies Eq.~(\ref{eq:djdt}) as shown by Littlejohn \cite{Littlejohn83}. Equation~(\ref{eq:dvdx_noncan}) is very useful, because it can be easily converted to other velocity coordinates that are defined locally in position space --- such as the pitches $\lambda = v_\parallel/v$ and $\Lambda = \mu B_0/E$ --- since their Jacobian factors can be derived analytically (\ref{apdx:bin}) \cite{Moseev19}.

Both (i) densities and (ii) histograms are {\it true} distributions in the sense that, when integrated over a phase space volume element $\Delta\V$,
\begin{equation}
\int_{\Delta\V}{\rm d}{\bm Z}\; h({\bm Z}) = \int_{\Delta\V}{\rm d}^3{\bm v}{\rm d}^3{\bm x}\;f = \N(\Delta\V),
\label{eq:n}
\end{equation}

\noindent they yield the true number $\N$ of objects contained inside that volume element. Functions or data that do not satisfy this condition {\it for the specified kind of objects} constitute the third type in our classification scheme:
\begin{enumerate}
\item[(iii)]  The symbol $G({\bm Z})$ represents {\it quasi-distributions}, which do not yield the true number of objects by mere volume integration in the configuration at hand, but require more complex transformations.
\end{enumerate}

\noindent Quasi-distributions often arise in experimental or modeling work. A typical example in experimental measurements is the line-of-sight density, where the transformation $G \rightarrow f$ takes the form of an inverse problem.

In fact, since the magnetic configuration in an experiment is rarely known accurately, any experimentally measured particle statistics should be treated as quasi-distributions when used as input for modeling distributions in a numerically constructed MHD equilibrium. Conversely, it should be kept in mind that the spatial distribution of GCs differs from that of physical particles since the latter is broadened by gyration around the GCs. This means that, strictly speaking, the GC distributions that we deal with in this work constitute quasi-distributions from the viewpoint of physical particles. Even if our GC phase space density function $f$ is normalized to give the correct number of physical particles in the entire plasma volume, $\N(\V) = \N_{\rm phys}(\V)$, the values of $\N(\Delta\V)$ and $\N_{\rm phys}(\Delta\V)$ given by Eq.~(\ref{eq:n}) for an arbitrary subvolume $\Delta\V$ may differ in the case of fast ions with a steep density gradient.

In general, the process of ``modeling'' essentially consists of choosing a physically meaningful transformation $G \rightarrow f$. We will return to this topic in Sections~\ref{sec:mdl} and \ref{sec:viz}, where we apply our methods to model a distribution of fusion-born alpha particles, based on a quasi-distribution $G(\Psi_{\rm P})$ that represents their birth profile.

Finally, we emphasize that when a GC orbit distribution function is given in the form $f_{\rm orb}(E,\mu,P_\zeta;\sigma)$, or an equivalent set of CoM coordinates, the Jacobian $\J_{\rm CoM}$ must be accompanied by instructions concerning how to deal with the index $\sigma$ that specifies the sign of the parallel velocity. First, it must be clarified whether $\sigma$ is the sign relative to the magnetic field or the plasma current if their directions differ. Second, it must be clarified how particles trapped in a magnetic mirror are to be counted, because $\sigma$ is redundant for such trapped particle orbits. Depending on the choice made, the integral in Eq.~(\ref{eq:n}) can be written in at least three ways:
\begin{subequations}
\begin{align}
\N =& \sum\limits_{\sigma = \pm 1} \int{\rm d}E{\rm d}\mu{\rm d}P_\zeta \underbrace{\J_{\rm CoM}^{\rm t/p}({\rm trap./pass.}) f_{\rm orb}}\limits_{h_{\rm orb}^{\rm t/p}(E,\mu,P_\zeta;\sigma)},
\label{eq:n_tp}
\\
=& \sum\limits_{\sigma_{\rm pass} = 0,\pm 1} \int{\rm d}E{\rm d}\mu{\rm d}P_\zeta \underbrace{\J_{\rm CoM} f_{\rm orb}}\limits_{h_{\rm orb}(E,\mu,P_\zeta;\sigma_{\rm pass})},
\label{eq:n_full}
\\
=& \sum\limits_{\sigma_{\rm HFS} = \pm 1} \sum\limits_{\sigma_{\rm LFS} = \pm 1} \int{\rm d}E{\rm d}\mu{\rm d}P_\zeta \underbrace{\frac{\J_{\rm CoM}}{2} f_{\rm orb}}\limits_{h_{\rm orb}(E,\mu,P_\zeta;\sigma)}.
\label{eq:n_double}
\end{align}
\label{eq:n_sigma}
\end{subequations}

\noindent In the first case (\ref{eq:n_tp}), the summation over $\sigma$ counts each trapped orbit twice, which means that the histogram $h^{\rm t/p}_{\rm orb}$ has two sets of identical entries each representing {\it half} of the particles in the domain of trapped orbits. In the second case (\ref{eq:n_full}), the summation index $\sigma_{\rm pass} = \{0,\pm 1\}$ is three-valued, identifying trapped ($0$), co- and counter-passing orbits ($\pm 1$), so that $h_{\rm orb}$ represents the {\it full} number of particles in all regions of GC phase space without duplicate entries. In mathematical form:
\begin{gather}
h_{\rm orb}(\sigma_{\rm pass} = 0) = 2\times h_{\rm orb}^{\rm t/p}({\rm trap.}),
\\
\J_{\rm CoM} = \J_{\rm CoM}^{\rm t/p}({\rm pass.}) = 2\times \J_{\rm CoM}^{\rm t/p}({\rm trap.}).
\end{gather}

\noindent The values of the density function $f_{\rm orb}$ itself are, of course, independent of the convention used. The point we want to make is that it is crucial to ensure consistency between the summation over $\sigma$ and the Jacobian when counting marker particles by integrating $f_{\rm orb}$ in $(E,\mu,P_\zeta;\sigma)$ space as in Eq.~(\ref{eq:n_sigma}).

In this paper, we adopt the third option (\ref{eq:n_double}), with the single-valued constant Jacobian $\J_{\rm CoM}/2$. Our CoM phase space mesh defined in Section~\ref{sec:mesh} will double-count all orbits --- passing and trapped --- when they cross the plasma midplane: once on the high-field side (HFS) and once on the low-field side (LFS). The factor $2$ dividing the Jacobian in Eq.~(\ref{eq:n_double}) compensates this.

\section{Geometry, coordinates and normalization}
\label{sec:geom}

\begin{figure}
[tb]
\centering
\includegraphics[width=0.47\textwidth]{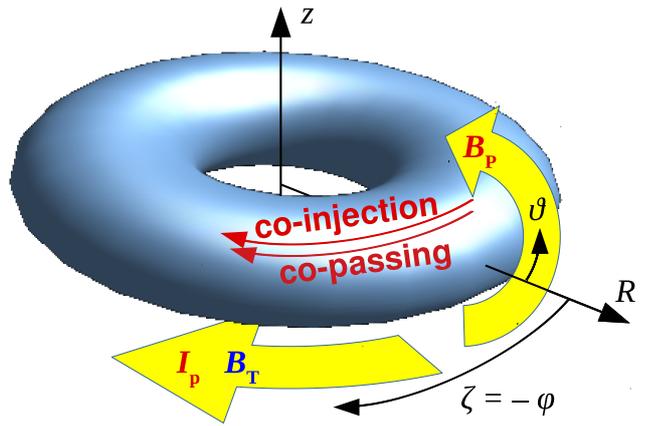}
\caption{Toroidal geometry and coordinates. Right-handed cylinder coordinates $(R,z,\zeta)$ and the poloidal angle $\vartheta$ are indicated. The yellow arrows show the orientation of the toroidal magnetic field ${\bm B}_{\rm T}$, the plasma current $I_{\rm p}$ and the associated poloidal magnetic field ${\bm B}_{\rm P}$ in our working example, which corresponds to the situation in typical JET and JT-60U tokamak plasmas. The co-/counter injection of beams is defined relative to the plasma current, as is the direction of co-/counter passing particles. Definitions in the literature vary, but in the present case there is no risk of confusion since $I_{\rm p}$ and ${\bm B}_{\rm T}$ are aligned.}
\label{fig:geometry}%
\end{figure}

\paragraph{System geometry} The magnetic field vector is written
\begin{align}
{\bm B} = \nablab\times{\bm A} =& \nablab\zeta \times \nablab\Psi_{\rm P} + B_\zeta\nablab\zeta \\
=& \nablab\zeta \times \nablab\Psi_{\rm P} + \nablab\Psi\times\nablab\vartheta,
\end{align}

\noindent where $2\pi\Psi = \frac{1}{2\pi}\int{\rm d}\Psi{\rm d}\vartheta{\rm d}\zeta\; \J_{\rm f}^{\bm x} {\bm B}\cdot\nablab\zeta$ is the toroidal flux and $2\pi\Psi_{\rm P} = -2\pi A_\zeta = \frac{1}{2\pi}\int{\rm d}\Psi{\rm d}\vartheta{\rm d}\zeta\; \J_{\rm f}^{\bm x} {\bm B}\cdot\nablab\vartheta$ the poloidal flux. These fluxes are related via the tokamak safety factor $q = {\rm d}\Psi/{\rm d}\Psi_{\rm P}$ measuring the mean magnetic field line pitch. $\J_{\rm f}^{\bm x} = [\nablab\Psi\cdot(\nablab\vartheta\times\nablab\zeta)]^{-1} = 1/B^\zeta$ is the Jacobian for the transformation from Cartesian position coordinates ${\bm x}$ to the toroidal flux coordinates $(\Psi,\vartheta,\zeta)$, where $\vartheta$ is a poloidal angle and $\zeta = -\varphi$ is the geometric toroidal angle, whose orientations are indicated in Fig.~\ref{fig:geometry}. Also shown in Fig.~\ref{fig:geometry} are the axes of the right-handed cylinder coordinate system $(R,z,\zeta)$, with major radius $R = \sqrt{x^2 + y^2}$ and vertical coordinate $z$.

\begin{figure}
[tbp]
\centering
\includegraphics[width=0.47\textwidth]{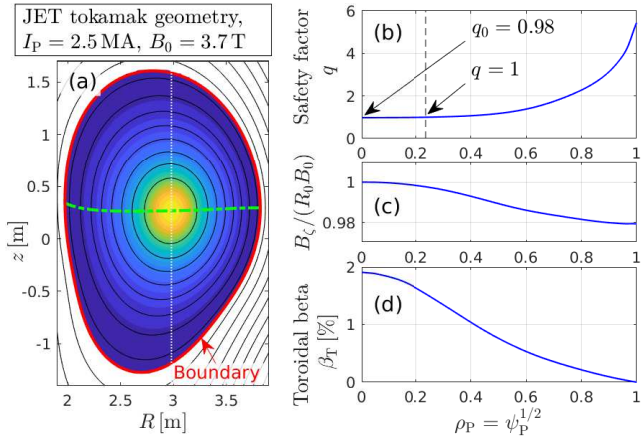}
\caption{Our working example takes the basic plasma geometry from JET \protect\cite{Nocente20,Kazakov21}, with plasma current $I_{\rm P} = 2.5\,{\rm MA}$, on-axis magnetic field $B_0 = 3.7\,{\rm T}$, and magnetic axis location $(R_0,z_0) = (2.98\,{\rm m},0.27\,{\rm m})$. Panel (a) shows contours of the poloidal flux $\Psi_{\rm P}(R,Z)$ (black) and scalar pressure $P(R,Z)$ (color), as well as the midplane (green, Eq.~(\protect\ref{eq:midplane})). The X-point has been removed and the plasma boundary (red) is located slightly inside of would have been the last closed flux surface. Panels (b)--(d) show the radial profiles of the safety factor $q = {\rm d}\Psi/{\rm d}\Psi_{\rm P}$, the covariant toroidal field component $B_\zeta$ and the toroidal beta $\beta_{\rm T} = 2\mu_0 P/B_0^2$ (with $\mu_0 = 4\pi\times 10^{-7}\,{\rm H/m}$) as functions of $\rho_{\rm P}$ defined in Eq.~(\ref{eq:rho}).}
\label{fig:tok}%
\end{figure}

The magnetic flux surface geometry and profiles for the MHD equilibrium that we use as a working example in this paper is shown in Fig.~\ref{fig:tok}. This plasma is partly based on recent experiments with central radio-frequency (RF) heating using a 3-ion scheme \cite{Nocente20, Kazakov21} performed at the Joint European Torus (JET). The magnetic X-point has been removed and the profile of the safety factor $q$ is chosen to increase monotonically from values near unity in the core ($q_0 = 0.98$ on axis) to $q_a = 5.44$ at the boundary. This simulation scenario is also used in ongoing studies of MHD instabilities and fast ion physics \cite{Bierwage22b}. The $q$ profile in the above-mentioned experiments varied dynamically and is likely to differ from ours at most times.

Besides $R$ and $z$, we will also use the auxiliary coordinates $X = R - R_0$ and $Y = z - z_0$ measuring horizontal and vertical positions relative to the magnetic axis, which is located here at $(R_0,z_0) = (2.98\,{\rm m},0.27\,{\rm m})$.

\paragraph{Magnetic flux labels} The normalized toroidal and poloidal fluxes ($\psi$, $\psi_{\rm P}$) or their square roots ($\rho_{\rm T}$, $\rho_{\rm P}$) can serve as convenient minor radial coordinates:
\begin{equation}
\rho_{\rm T}^2 = \psi = \frac{\int_0^{V(\Psi)}{\rm d}V/\J_{\rm f}^{\bm x}}{\int_0^{V(\Psi_{\rm a})}{\rm d}V/\J_{\rm f}^{\bm x}}, \quad
\rho_{\rm P}^2 = \psi_{\rm P} = \frac{\Psi_{\rm P} - \Psi_{\rm P,0}}{\Psi_{{\rm P},a} - \Psi_{\rm P,0}},
\label{eq:rho}
\end{equation}

\noindent where $V(\Psi)$ is the volume within the flux surface labeled $\Psi$. $\Psi_{\rm P,0}$ and $\Psi_{{\rm P},a}$ are the values of the poloidal flux at the magnetic axis ($r=0$) and boundary ($r=a$). For the toroidal flux we have $\Psi_a - \Psi_0 = \int_{\Psi_{{\rm P},0}}^{\Psi_{{\rm P},a}} {\rm d}\Psi_{\rm P}\, q$. In some occasions, we use a volume-averaged minor radius $0 \leq r(\Psi) \leq a$, which is defined here as
\begin{equation}
r^2 = \frac{R_0}{\pi} \int_0^{V(r)} \frac{{\rm d}V}{\J_{\rm f}^{\bm x} B_\zeta} = \frac{R_0}{\pi} \int_0^{\Psi(r)} \frac{{\rm d}\Psi}{B_\zeta} \approx a^2\psi.
\label{eq:r}
\end{equation}

\noindent Our $r$ reduces the geometric minor radius in the limit of a cylindrical plasma with circular cross-section. For $B_\zeta \approx R_0 B_0$, we have $r/a \approx \rho_{\rm T}$, and $a \approx [2(\Psi_a-\Psi_0)/B_0]^{1/2}$ is the minor radius of the plasma boundary.

\paragraph{Particle and orbit coordinates} In the guiding center (GC) description, the velocity space consists of the ignorable gyrophase $\xi$, the perpendicular particle velocity $v_\perp$, and the parallel GC velocity $v_\parallel = {\bm v}_{\rm gc}\cdot\hat{\bm b}$ relative to the magnetic field vector ${\bm B}$ with unit vector $\hat{\bm b} \equiv {\bm B}/B$ and magnitude $B = |{\bm B}|$. The canonical momentum can be written in the form ${\bm P} = {\bm A}/B_0 + \Omega_0^{-1}{\bm v}$ for a particle with gyrofrequency $\Omega_0 = QeB_0/M$, charge $Qe$ and mass $M$. In GC coordinates, its covariant toroidal component,
\begin{equation}
P_\zeta = {\bm P}_{\rm gc}\cdot\partial_\zeta{\bm x} = -\frac{\Psi_{\rm P}}{B_0} + \frac{v_\parallel}{\Omega_0} \frac{B_\zeta}{B},
\label{eq:pzeta}
\end{equation}

\noindent is called the canonical toroidal angular momentum and is chosen here to have units of area per radian, $(2\pi)^{-1}[{\rm m}^2]$. Together with the lowest-order magnetic moment $\mu$ and the signed kinetic energy $\enr = \sigma_{\rm ref}E$,
\begin{equation}
\mu = \frac{M v_\perp^2}{2 B}, \quad \enr = \sigma_{\rm ref}\frac{M}{2} v^2, \quad \sigma = \frac{v_\parallel J_\parallel}{|v_\parallel J_\parallel|},
\label{eq:def_esgn_mu}
\end{equation}

\noindent this yields a set of coordinates $(\enr,\mu,P_\zeta)$ that belongs to the class of constants of motion (CoM), which are conserved by our equations of motion (presented in Section~\ref{sec:eom}) when they are solved in a time-independent field ${\bm B}$ that is symmetric along $\zeta$ (or ``axisymmetric'' around $\hat{\bm e}_z$), satisfying $\partial_\zeta B_\zeta = \partial_\zeta\Psi_{\rm P} = 0$.

The sign $\sigma$ of the parallel velocity is measured relative to the parallel current density $J_\parallel = {\bm J}\cdot{\bm B}/B$. While redundant or unnecessary for orbits trapped in a magnetic mirror (both signs are present), $\sigma$ identifies whether a passing orbit with coordinates $(E,\mu,P_\zeta)$ is co- or counter-going. In mathematical treatments, $\sigma$ usually appears as a separate index, $(E,\mu,P_\zeta;\sigma)$, but this is inconvenient for numerical representations. In our orbit database, we define $\sigma_{\rm ref}$ at a certain reference point; namely, the starting point used for the orbit calculation. This or any equivalent convention allows us to attach $\sigma_{\rm ref}$ to $E$ as in Eq.~(\ref{eq:def_esgn_mu}) and obtain a compact representation where all CoM arrays in the computer code have the same dimensionality: three.

It is also useful to define several different coordinates measuring the velocity pitch:
\begin{equation}
\Lambda \equiv \frac{\mu B_0}{E} = \frac{\cos^2\alpha}{\hat{B}}, \quad \sin\alpha = \lambda \equiv \frac{v_\parallel}{v} = \sigma_B\sqrt{1 - \Lambda\hat{B}},
\end{equation}

\noindent with $\hat{B} = B/B_0 \approx R_0/R$ and $\sigma_B \equiv v_\parallel/|v_\parallel|$. Each pitch coordinate has its advantages in practical applications:
\begin{itemize}
\item The coordinate $\lambda = v_\parallel/v$ is widely used together with $E$ or velocity $v = \sqrt{2E/M}$. The local velocity space Jacobian of this set is a CoM ($J_{E\lambda} \propto \sqrt{E}$, Eq.~(\ref{eq:jv2})), and the Fokker-Planck equation can be integrated analytically in these coordinates \cite{Weiland18}.

\item The coordinate $\alpha$ is a true angle, so its value measures the velocity pitch in a very intuitive form. Moreover, the domains of co- and counter-passing particles are enlarged along the $\alpha$-axis, making this coordinate a good choice for sampling the velocity distribution of tangentially injected beam ions \cite{Bierwage18} and distributions produced by central RF heating using a 3-ion scheme \cite{Nocente20, Kazakov21, Stancar21}.

\item  The coordinate $\Lambda = \hat{B}^{-1} \cos^2\alpha$ has the advantage of being a CoM and that (in contrast to $\mu$) its upper limit is independent of energy: $0 \leq \Lambda \leq \hat{B}_{\rm min}^{-1}$, where $B_{\rm min} = {\rm min}\{\hat{B}\}$ is the minimal field strength in the considered domain. This makes $\Lambda$ a convenient choice for defining the lines along which to divide the phase space as will be done in Section~\ref{sec:mesh}.
\end{itemize}

\begin{figure}
[tbp]
\centering
\includegraphics[width=0.47\textwidth]{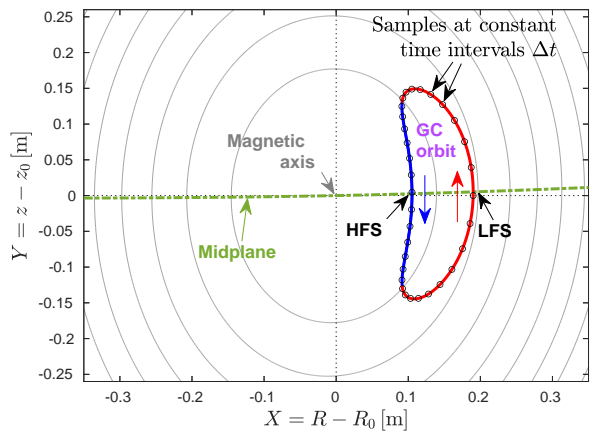}
\caption{GC orbit of a core-localized trapped alpha particle with kinetic energy $E = 1.5\,{\rm MeV}$ and pitch angle $\alpha = 0.056\pi$. This figure shows the inner region of the poloidal plasma cross-section of Fig.~\protect\ref{fig:tok}(a). The gray contours are uniformly spaced in poloidal magnetic flux $\Psi_{\rm P}$. The orbit is sampled uniformly in time (black circles). The GC is co-moving ($v_\parallel \geq 0$) in the red and counter-moving ($v_\parallel < 0$) in the blue portion of the orbit contour. The high- and low-field-side intersections of the orbit with the midplane are labeled HFS, LFS.}
\label{fig:midplane}%
\end{figure}

Unlike the conserved quantities $E$, $\mu$ and $\Lambda$, the pitch coordinates $\alpha$ and $\lambda$ vary along a GC orbit. The latter can be turned into orbit coordinates (i.e., CoM), $\alpha_{\rm ref}$ and $\lambda_{\rm ref}$, by evaluating them at a well-defined reference point, as we did earlier for the sign $\sigma_{\rm ref}$.

\paragraph{Reference points for orbit coordinates} In cases where we import computed particle distributions (blue workflow in Fig.~\ref{fig:vstart}), the initial reference point is taken to be the position of an imported marker particle, which may be located anywhere inside the plasma or in the surrounding vacuum.

In cases where we need to construct a new CoM mesh (red workflow in Fig.~\ref{fig:vstart}) the reference points lie in the midplane, which (at least in a usual tokamak plasma) contains all O-type stagnation points and is defined as the curve $z_{\rm m}(X)$ along which the magnetic field is perpendicular to its gradient,
\begin{equation}
z_{\rm m}(R)\; :\; {\bm B}\cdot\nablab B = 0,
\label{eq:midplane}
\end{equation}

\noindent The midplane is shown as a dash-dotted green line in Fig.~\ref{fig:tok}. It contains the magnetic axis, $(R,z) = (R_0,z_0)$, but deviates from the horizontal plane $z = z_0$ in up-down asymmetric plasmas like ours. Every GC orbit crosses the midplane twice, once in a region with higher field strength and once in a region with lower field strength. Figure~\ref{fig:midplane} shows an example, where the orbit's high- and low-field side crossings of the midplane are labeled ``HFS'' and ``LFS'', respectively.

\paragraph{Normalization} Spatial positions and lengths are given in meters unless stated otherwise. Velocities are normalized by a reference value $v_0$. Energy is normalized by $M v_0^2$ (for each particle species) and the magnetic field by its on-axis value $B_0$ (here $3.7\,{\rm Tesla}$):
\begin{equation}
\hat{v} = \frac{v}{v_0}, \quad \hat{E} = \frac{E}{Mv_0^2} = \frac{\hat{v}^2}{2}, \quad \hat{\mu} = \frac{\mu B_0}{Mv_0^2} = \frac{\hat{v}_\perp^2}{2\hat{B}}.
\end{equation}

\section{CoM mesh and drift orbit types}
\label{sec:mesh}

\begin{figure}
[tb]
\centering
\includegraphics[width=0.47\textwidth]{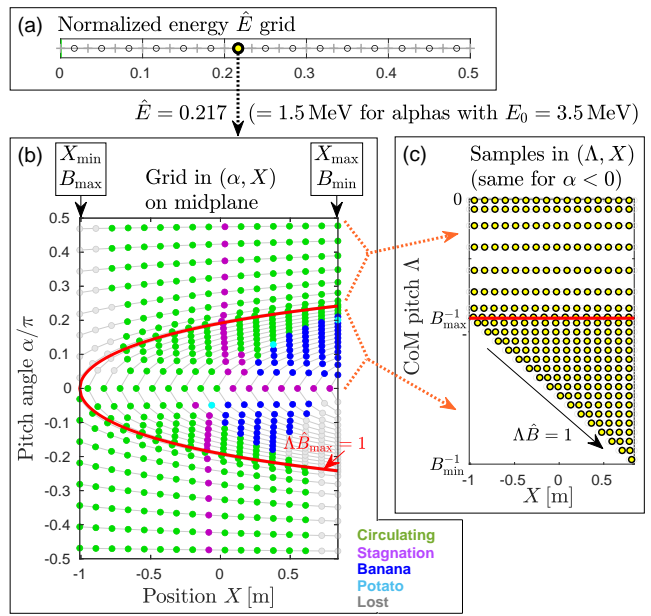}
\caption{Method for sampling the GC orbit space in {\tt VisualStart}. Panel (a) shows the grid in normalized kinetic energy $\hat{E} = \hat{v}^2/2$, here chosen to be uniformly spaced in $\hat{v}^2$. Panel (b) shows the grid in pitch angle-position space $(\alpha,X)$ at the midplane of Fig.~\protect\ref{fig:tok}(a). The color of each grid point in (b) identifies the type of an orbit for a given energy, here $\hat{E} = 0.217$. These grid points lie along lines of $\Lambda = {\rm const}$., which appear parabolic in the $(\alpha,X)$-plane of panel (b) and straight in panel (c). The black circles in (c) indicate the locations of the samples taken halfway between grid points. The bold red line indicates the contour where $\Lambda\hat{B}_{\rm max} = 1$. Each orbit is effectively sampled twice: trapped orbits appear above and below the $\alpha = 0$ line, and circulating orbits on the left and right of the stagnation points. Thus, except very close to a stagnation point, passing and trapped orbits can be treated equally, because both are effectively double-counted in this mesh.}
\label{fig:mesh}%
\end{figure}

In this section, we describe how we sample the GC orbit space. The example in Fig.~\ref{fig:mesh} is used for illustration. A low resolution is chosen in order to make all grid points clearly visible. All coordinates shown in Fig.~\ref{fig:mesh} are evaluated at the height $z_{\rm m}(X)$ of the midplane as defined in Eq.~(\ref{eq:midplane}). This means that all coordinates appearing in this section are constants of motion (CoM). The set $(\hat{E},\alpha,X)$ used in panels (a) and (b) yields particularly compact and, in our opinion, intuitive plots. Another view of pitch-position space in $(\Lambda,X)$ coordinates is shown in panel (c). The plasma boundary is also taken to be the loss boundary for the GC orbits, so we do not include the vacuum here and our midplane mesh covers only the width of the plasma itself: $-1\,{\rm m} \lesssim X \lesssim 0.84\,{\rm m}$.

The reference velocity used for normalization is chosen to be $v_0 = 1.3\times 10^7\,{\rm m/s}$, which corresponds to $E_0 = M v_0^2/2 \approx 3.5\,{\rm MeV}$ for fusion-born alpha particles ($^4_2{\rm He}^{+2}$) or $1.75\,{\rm MeV}$ for deuterons ($^2_1{\rm H}^+$). Both species have the same charge-to-mass ratio $Qe/M$ and their characteristic gyroradius in this example is $\rho_0 = v_0/\Omega_0 \approx 0.07\,{\rm m} \approx 0.02 \times R_0$. We consider the full energy range $0 \leq \hat{E} \leq \hat{E}_0$ with $\hat{E}_0 = 0.5$ and define a mesh consisting of $N^{\rm (g)}_E$ grid points that are the boundaries of $N_E = N^{\rm (g)}_E - 1$ cells. The value of energy at the center of cell $i$ is denoted $\hat{E}_i$. The example in Fig.~\ref{fig:mesh}(a) has $N_E = 15$ cells with equal sizes $\Delta\hat{E}_i = \hat{E}_0/N_E = {\rm const}$.

The remaining phase space is divided along the coordinate lines of the conserved pitch $\Lambda$. These lines appear parabolic in the $(\alpha,X)$-plane of panel (b) and horizontal in the $(\Lambda,X)$-plane of panel (c). Two different procedures are used in the domains above and below the red line in Fig.~\ref{fig:mesh}, which represents $\Lambda\hat{B}_{\rm max} = 1$:
\begin{itemize}
\item  $0 < \Lambda < \hat{B}_{\rm max}^{-1}$: At $X = X_{\rm min}$, namely the left vertical axis of panel (b), we define a grid in the pitch angle coordinate $-\pi/2 \leq \alpha \leq \pi/2$. In the present example, it consists of $N_\alpha = 16$ equally sized cells with $\Delta\alpha_j = \pi/N_\alpha = {\rm const}$. The circles at $X = X_{\rm min}$ in panel (b) indicate the cell centers $\alpha_j$ with $j = 1...N_\alpha$. These points $(\alpha_j,X_{\rm min})$ are the starting points for lines $\Lambda = \cos^2\alpha/\hat{B} = {\rm const}$.\ plotted gray in Fig.~\ref{fig:mesh}(b). Along each of these lines, we define a grid in the radial coordinate $X$. In the present example, it consists of $N_X = 16$ cells with $\Delta X_k = (X_{\rm max}-X_{\rm min})/N_X = {\rm const}$. The upper part of panel (c) shows the resulting samples $(\Lambda_j,X_k)$ at the cell centers in the $\alpha > 0$ portion of panel (b).

\item  $B_{\rm max}^{-1} < \Lambda < \hat{B}_{\rm min}^{-1}$: At $\alpha = 0$, namely the horizontal axis at the center of panel (b), we define a grid in the radial coordinate $X$. Our example has $N_X = 16$ equally sized cells. The circles at $\alpha = 0$ indicate the cell centers $X_j$ with $j = 1...N_X$, which form a diagonal ($\Lambda\hat{B} = 1$) in the lower part of panel (c). They are the origins of grid lines $\Lambda^{\rm (g)}_j = {\rm const}$., which are plotted gray in panel (b). Here, we choose to have an approximately uniform sample density. For this purpose, we define an auxiliary coordinate $0 \leq d_j \leq L_j$ that measures the distance along a grid line $\Lambda^{\rm (g)}_j$ in the $(\hat{\alpha},\hat{X})$-plane, with $\hat{\alpha} = \alpha/(2\pi)$ and $\hat{X} = X/(X_{\rm max} - X_{\rm min})$. Along each grid line, we create $N_j$ cells with roughly equal sizes $\Delta d_{j,k} = L_j/N_j \approx {\rm const}$. The number of cells satisfies $2 \leq N_j \leq 2N_X$ and the cell index $1 \leq k \leq N_j$ covers positive and negative pitches, so there is at least 1 cell and at most $N_X$ cells in each domain, $\alpha \gtrless 0$, respectively. The lower part of panel (c) shows the resulting samples $(\Lambda_j,X_k)$ at the cell centers in the $\alpha > 0$ portion of panel (b).
\end{itemize}

\noindent The above procedure samples the phase space of confined GC orbits twice, once on the LFS and once on the HFS. This ensures full coverage of the phase space and eliminates the need to distinguish between orbit types inside the mesh-cutting algorithm. The double-counting will be corrected by a factor $1/2$ in the volume elements derived in Section~\ref{sec:weight} below.

The color of each sample in Fig.~\ref{fig:mesh}(b) identifies the type of an orbit for a given energy. Here, we have chosen $\hat{E}_{i=7} = 0.217$. In the case of alpha particles with $E_0 = Mv_0^2/2 = 3.5\,{\rm MeV}$, this corresponds to an energy of $1.5\,{\rm MeV}$. Orbits are divided into two groups: orbits that are trapped in a magnetic mirror, and passing orbits (co- and counter-current). Trapped orbits contain a point where $v_\parallel$ changes sign, while passing orbits do not. Each group is further divided into two sub-categories as defined in Table~\ref{tab:orbits}. This orbit classification scheme is relatively simple; see Ref.~\cite{Rome79} for more elaborate discussions. Note that for every tuple $(E,\Lambda)$, the midplane contains a pair of O-type stagnation points, where co- or counter-passing stagnation orbits are point-like in the poloidal plane. Since these O-type stagnation points alone are insignificant, we have chosen to expand the class of ``stagnation orbits'' to include all passing orbits that do not encircle the magnetic axis.

\begin{table}[tb]
\begin{tabular}{c|c|c}
\hline\hline & \multicolumn{2}{c}{Encircle magnetic axis at $(X,Y) = (0,0)$} \\
& Yes & No \\
\hline Trapped: & Potato orbits & Banana orbits \\
Passing: & Circulating orbits & Stagnation orbits \\
\hline\hline
\end{tabular}
\caption{Orbit classification.}
\label{tab:orbits}
\end{table}

The classes of potato and stagnation orbits arise from magnetic drifts. The magnitude of the magnetic drift velocity ${\bm v}_{\rm dB}$ (see Eq.~(\protect\ref{eq:v}) below) is proportional to the inverse aspect ratio $a/R_0$ and the kinetic energy. Furthermore, its $(R,z)$-component is inversely proportional to the poloidal field $B_{\rm P} \propto I_{\rm p}$, which tends to divert it into the toroidal direction. This means that magnetic drifts tend to be large for energetic particles in tokamak plasmas with relatively low plasma current and in compact tori. In such cases, the overall fraction of potato and stagnation orbits can be significant.

The mesh in Fig.~\ref{fig:mesh}(b) is rather sparse, and our automatic classification algorithm identified only 3 samples as potato orbits (see Fig.~\ref{fig:com_increment_40}(a) for a denser mesh and higher energy). The $\alpha > 0$ legs of potato orbits are found around the upper rim of the trapped-passing boundary on the $X > 0$ side of Fig.~\ref{fig:mesh}(b). On the $X < 0$ side, they are located slightly below the $\alpha = 0$ line.

\begin{figure}
[tb]
\centering
\includegraphics[width=0.47\textwidth]{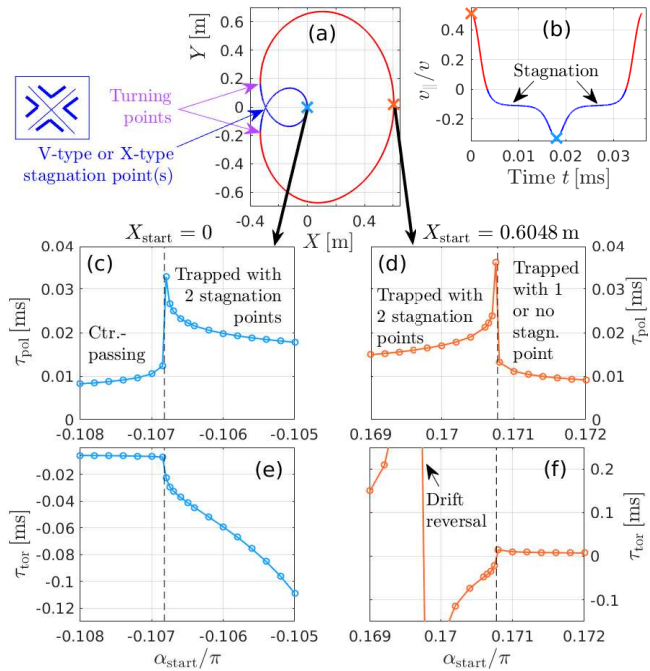}
\caption{Properties of GC orbits near a trapped-passing (t-p) boundary. Panel (a) shows the poloidal orbit contour of a trapped alpha particle with $E = 3.5\,{\rm MeV}$ located very close to the t-p boundary, which is a topological discontinuity occurring at the X-point of the orbit's counter-going leg ($v_\parallel < 0$, blue). The box on the top left illustrates schematically the topological transition between two pairs of V-type stagnation points via an X-point at the t-p boundary. The turning points, where $v_\parallel$ changes sign, are also indicated in (a). The time trace of the pitch $v_\parallel / v = \sin\alpha$ is shown in panel (b) with time $t$ in milliseconds. Panels (c)--(f) show the pitch angle dependence of the poloidal transit time $\tau_{\rm pol}$ (c,d) and the toroidal transit time $\tau_{\rm tor}$ (e,f) across the t-p boundary (vertical dashed line). The starting point $(E,\alpha_{\rm start},X_{\rm start})$ lies on the midplane ($Y \approx 0$) as indicated by the crosses in panel (a): $X_{\rm start} = 0$ is on the inner leg with $\alpha < 0$ for (c,e), and $X_{\rm start} \approx 0.6\,{\rm m}$ is on the outer leg with $\alpha > 0$ for (d,f). The poloidal transit time $\tau_{\rm pol}$ exhibits singular behavior at the t-p boundary. Meanwhile, the apparent discontinuity merely reflects the partition of a phase space volume element between two orbits upon crossing the t-p boundary.}
\label{fig:tp-boundary}%
\end{figure}

It is important to note that the trapped-passing (t-p) boundary on the $X < 0$ side also lies {\it below} the $\alpha = 0$ line. This is due to the magnetic drifts and has the implication that the turning points --- where the magnetic mirror force causes a sign reversal in $v_\parallel$ --- are near but not identical to the V-type stagnation points of trapped particle orbits.\footnote{The discussion below Eq.~(19) in Ref.~\protect\cite{Porcelli94} is related to this.}
The example in Fig.~\protect\ref{fig:tp-boundary}(a) illustrates this for the case of an orbit very close to a t-p boundary.

The t-p boundary refers to a separatrix in the orbit topology; i.e., it marks the topological transition of two V-type stagnation points via an X-type stagnation point as illustrated schematically in the box at the top left of Fig.~\ref{fig:tp-boundary}. The marginally trapped orbit in panel (a) has two V-type stagnation points that nearly form an X. Across the t-p boundary, this orbit decomposes into a counter-passing orbit and a trapped orbit, each with a single V-type stagnation point that turns into a smooth part of the orbit contour as one departs from the t-p boundary.

We use the word ``stagnation'' to indicate that GCs spend a relatively large amount of time in a relatively small region of the poloidal plane. Around O-type stagnation points, this is the case for the entire orbit. In the case of X- and V-type stagnation points as in Fig.~\ref{fig:tp-boundary}, GCs spend a large amount of time in a small portion of the orbit contour. Spatial stagnation also means that $v_\parallel/v = \sigma_B\sqrt{1 - \Lambda\hat{B}}$ stagnates as shown in panel (b).

The pitch angle scans of the poloidal and toroidal transit times in panels (c)--(f) of Fig.~\ref{fig:tp-boundary} show true singularities in $\tau_{\rm pol}$ and $\tau_{\rm tor}$, and a topological discontinuity (see also Fig.~3 of Ref.~\cite{Zhao97}). The role of the transit time singularities will be discussed later in Section~\ref{sec:weight}. The discontinuity is a matter of perspective, as it merely reflects how a larger volume element of GC phase space is divided into two components represented by a pair of orbits that were unified on the other side of the t-p boundary. In some sense, the transition is smooth if one views the disconnected orbits as a pair.

Finally, we note that $v_\parallel = 0$ constitutes a degeneracy in CoM space, since $\Lambda$ and $P_\zeta(X)$ cannot be varied independently for fixed $E$ at that location. This is why, geometrically, the parabolic $\Lambda = {\rm const}$.\ curves have their pole at that location in Fig.~\ref{fig:mesh}. Since we are defining our CoM mesh in the midplane, we are able to avoid these degenerate $v_\parallel = 0$ points entirely by locating them along the $X_k$ cell boundary on $(E,\Lambda) = {\rm const}$.\ lines. All cell centers are, thus, ensured to have $v_\parallel \neq 0$.

\section{Equations of motion}
\label{sec:eom}

As indicated by the red and blue arrows between Steps 2 \& 3 in our workflow Fig.~\ref{fig:vstart}, the GC phase space samples that are provided by
\begin{enumerate}
\item[(a)] the mesh constructed in Section~\ref{sec:mesh}, or
\item[(b)] particle codes like {\tt OFMC} \cite{Tani81,Tani08} and {\tt MEGA} \cite{Todo98,Todo05} (when operated in full-$f$ mode \cite{Bierwage13a,Todo14b})
\end{enumerate}

\noindent serve as initial conditions for the GC equations of motion, which we solve once around each orbit.

{\tt VisualStart} solves the same equations with the same numerical scheme as its current target code {\tt MEGA}. Following the formalism introduced by Littlejohn \cite{Littlejohn83} and reviewed by Cary \& Brizard \cite{Cary09}, we use the effective electromagnetic fields ${\bm B}^* \equiv \nablab\times{\bm A}^*$ and ${\bm E}^* \equiv -\nablab\Phi^* - \partial_t{\bm A}^*$, with the potentials ${\bm A}^* \equiv {\bm A} + \frac{v_\parallel}{\Omega}{\bm B}$ and $Qe\Phi^* \equiv Qe\Phi + \mu B$. Here, we consider the motion of GCs in an unperturbed magnetic field and in the absence of any electric field ($\partial_t{\bm B} = \Phi = 0$), so the effective fields reduce to
\begin{equation}
{\bm B}^* = {\bm B} + \frac{v_\parallel}{\Omega} B \nablab\times\hat{\bm b}, \qquad {\bm E}^* = - \frac{\mu\nablab B}{Qe},
\label{eq:bestar}
\end{equation}

\noindent and the equations for the GC velocity ${\bm v}_{\rm gc} = \dot{\bm x}_{\rm gc}$ and its parallel acceleration become
\begin{align}
\dot{\bm x}_{\rm gc} =\;& v_\parallel\frac{{\bm B}^*}{B_\parallel^*} + \frac{{\bm E}^*\times\hat{\bm b}}{B_\parallel^*} = {\bm v}_\parallel^* + {\bm v}_{\rm dB}^*,
\label{eq:dxgc_dt}
\\
\dot{v}_\parallel =\;& \frac{\Omega}{B} \frac{{\bm B}^*\cdot{\bm E}^*}{B_\parallel^*}.
\label{eq:du_dt}
\end{align}

\noindent $\hat{\bm b} = {\bm B}/B$ is the unit vector along ${\bm B}$, and $B_\parallel^* \equiv {\bm B}^* \cdot\hat{\bm b}$ is Littlejohn's GC Jacobian for the noncanonical set of GC coordinates $(\mu,u,\xi,{\bm x}_{\rm gc})$ with $u \equiv \hat{\bm b}({\bm x}_{\rm gc})\cdot\dot{\bm x}_{\rm gc} \approx v_\parallel$ (see also the discussion on p.~704 of Ref.~\cite{Cary09}). The GC velocity is composed of a parallel component (modified by curvature effects) and magnetic drifts due to $\nablab B$ and curvature $\nablab\times\hat{\bm b}$:
\begin{equation}
{\bm v}_\parallel^* = v_\parallel\frac{\bm B}{B_\parallel^*}, \quad
{\bm v}_{\rm dB}^* = \frac{\mu}{Qe B_\parallel^*} \hat{\bm b}\times\nablab B + \frac{v_\parallel^2}{\Omega}\frac{B}{B_\parallel^*}\nablab\times\hat{\bm b}.
\label{eq:v}
\end{equation}

\noindent In the normalization of {\tt VisualStart} (Section~\ref{sec:mesh}):
\begin{equation}
\hat{\bm v}_\parallel^* = \hat{v}_\parallel\frac{\hat{\bm B}}{\hat{B}_\parallel^*}, \quad
\hat{\bm v}_{\rm dB}^* = \rho_0\frac{\hat{\mu}}{\hat{B}_\parallel^*} \hat{\bm b}\times\nablab \hat{B} + \rho_0\frac{\hat{v}_\parallel^2}{\hat{B}_\parallel^*} \nablab\times\hat{\bm b},
\label{eq:v_nrm}
\end{equation}

\noindent with $\rho_0 = v_0/\Omega_0$. The computed GC orbits are recorded in a database, following accuracy checks (\ref{apdx:num}).

The example in Fig.~\ref{fig:mesh}, gives a total of about 8000 orbit samples, 52\% of which lie inside the region bounded by the red curve ($\Lambda\hat{B}_{\rm max} \leq 1$) for the present setup. About 1100 orbits (14\%) are lost as they hit the (artificial) boundary of the simulation domain due to magnetic drifts. Using 3 processes on an 8th Generation Intel CORE i7-8565U processor ($1.80\,{\rm GHz}\times 8$), it took about 35 minutes to compute these 8000 orbits with our simple {\tt Matlab} code in which there is much room left for optimization. A proper orbit database for the entire plasma volume may require at least 100 times more samples, which would take 150 core-hours with the present implementation. An optimized code in a language such as {\tt Fortran}, {\tt C} or {\tt Julia} is expected to need only a small fraction of this time.\footnote{An experimental Julia version of our orbit solver runs 50 times faster.}
Making use of GPUs, the time required to compute such an orbit database should be reasonably short for routine work.

\section{Volume elements and weighting}
\label{sec:weight}

\subsection{Derivation}
\label{sec:weight_derive}

Having computed the GC orbits using the unperturbed equations of motion in Section~\ref{sec:eom} starting from the midplane mesh defined in Section~\ref{sec:mesh}, the next step is to represent the distribution function $f$ in discretized form using an ensemble of marker particles labeled $n = 1...N_{\rm mrk}$, which sample the orbits in our database. In other words, we seek the Klimontovich representation\footnote{\protect\ref{apdx:bin} describes in detail how we evaluate Eq.~(\protect\ref{eq:f_w}).}
\begin{align}
f({\bm Z}) \approx \sum_{n=1}^{N_{\rm mrk}} w_n\times\delta({\bm Z}_{{\rm gc},n} - {\bm Z}).
\label{eq:f_w}
\end{align}

\noindent At this point, ${\bm Z}_{{\rm gc},n}$ is the marker position in an arbitrary set of GC coordinates including ignorable angles, for instance $(v_\perp^2,v_\parallel,\xi,{\bm x}_{\rm gc})$. The Dirac $\delta$ distribution has units of inverse phase space volume $[{\rm m}^{-6}{\rm s}^{3}]$ and may be viewed as a proxy for any sort of particle shape factor needed to map the weights $w_n$ onto a discrete mesh. Each marker follows the trajectory of a GC as indicated by the subscript in ${\bm Z}_{{\rm gc},n}$. Its weight factor $w_n$ represents a certain number $[\Delta \N]_n$ of GCs of physical particles:
\begin{align}
w_n = [\Delta\N]_n =&\; \Delta\V_n \times \frac{\N(\Delta\V_n)}{\Delta\V_n}, \nonumber
\\
=&\; \Delta\V_n \times f.
\label{eq:w_f}
\end{align}

\noindent where $\Delta\V_n$ is a 6-D volume element of GC phase space in Cartesian coordinates that is attached to marker $n$.

Since we are considering only exact equilibrium distributions, the orbit time $\tau$ is an ignorable angle coordinate with period $\tau_{\rm pol}$ and all factors in Eq.~(\ref{eq:w_f}) are independent of the three angle coordinates $(\tau,\xi,\zeta)$. These distributions depend only on three CoM coordinates, whose values we will represent by cell indices $(i,j,k)$. When we load $N_\tau$ markers uniformly in time along an unperturbed GC orbit, neighboring markers are indistinguishable. Each marker carries precisely an $N_\tau$'s fraction of their orbit's volume $\Delta\V_{\rm orb}$ and weight $W$:
\begin{align}
\Delta\V_n \rightarrow \Delta\V_{ijkl} =& \left[\frac{\Delta\V_{\rm orb}}{N_\tau}\right]_{ijk} = \left[\frac{\Delta\V_{\rm orb} \Delta\tau_l}{\tau_{\rm pol}}\right]_{ijk},
\label{eq:vgc}
\\
w_n \rightarrow w_{ijkl} =& \left[\frac{W}{N_\tau}\right]_{ijk} \quad \text{for}\; l = 1,...,N_{\tau,ijk}.
\label{eq:w}
\end{align}

\noindent As indicated in Eq.~(\ref{eq:vgc}), the markers also represent equal fractions $\Delta\tau$ of an orbit's poloidal period $\tau_{\rm pol}$:
\begin{equation}
\Delta \tau_{ijkl} = \left[\frac{\tau_{\rm pol}}{N_\tau}\right]_{ijk}.
\label{eq:dt}
\end{equation}

\noindent Recalling from Eqs.~(\ref{eq:w_f}) that the volume element is eventually applied to a distribution function $f$, it is clear that Eq.~(\ref{eq:dt}) actually implies an orbit time average:
\begin{equation}
\sum_{l=1}^{N_{\tau,ijk}}\Delta\tau_{ijkl} f({\bm Z}_{{\rm gc},ijkl}) \approx \oint_{\tau_{\rm pol}}{\rm d}\tau\,f \equiv \tau_{\rm pol} \underbrace{\left<f\right>_{\rm orb}}\limits_{f_{\rm orb}}.
\label{eq:f_tavg}
\end{equation}

\noindent The average is trivial for true equilibrium distributions, which are independent of $\tau$, so that $f = f_{\rm orb}$. However, $f$ in Eq.~(\ref{eq:f_tavg}) may also be replaced by an arbitrary quasi-distribution $G$, so this equation also constitutes a recipe for transforming arbitrary models into true equilibria. In summary, Eq.~(\ref{eq:w_f}) becomes
\begin{equation}
w_{ijkl} = \left[\frac{W}{N_\tau}\right]_{ijk} = \left[\frac{\Delta \V_{\rm orb}}{N_\tau} \times f_{\rm orb}\right]_{ijk}.
\label{eq:w_f_com}
\end{equation}

In applications where we merely import particle data computed by another code (blue workflow in Fig.~\ref{fig:vstart}), the weights $w_l$ are provided with the data. Each sample is interpreted as the initial position of a new GC orbit, so the input weight is interpreted as the weight of an orbit: $[w_{\rm import}]_l \rightarrow W_{ijk}$. This step enforces the equilibrium constraint and the first equality of Eq.~(\ref{eq:w_f_com}) yields the marker weights. The volume element $\Delta\V$ and distribution function $f$ do not appear explicitly in such a case.

In applications where we construct a new distribution function from a model or where we reexpress a mesh-based distribution function in the orbit-based representation (red workflow in Fig.~\ref{fig:vstart}), it is necessary to determine the GC orbit volume element $[\Delta\V_{\rm orb}]_{ijk}$ for a given mesh in CoM space. In our case, the mesh has the form shown in Fig.~\ref{fig:mesh} and the associated volume elements $[\Delta\V_{\rm orb}]_{ijk}$ in units of $[{\rm m}^6 {\rm s}^{-3}]$ can be readily obtained when written in terms of the canonical action coordinates $(E,\mu,P_\zeta)$ associated with the ignorable angles $(\tau,\xi,\zeta)$, taking advantage of the fact that the Jacobian for a canonical transformation is constant in both space and time (see, for instance, Ref.~\cite{Porcelli94}).

With our definition $P_\zeta = -\Psi_{\rm P}/B_0 + v_\parallel B_\zeta/(\Omega_0 B)$, which has units of $(2\pi)^{-1}[{\rm m}^2]$, the transformation between Cartesian coordinates $({\bm v},{\bm x})$ for the phase space of physical particles and canonical coordinates $(E,\tau,\mu,\xi,P_\zeta,\zeta)$ for the phase space of GCs has the form
\begin{align}
{\rm d}^3{\bm v}{\rm d}^3{\bm x} = &\frac{1}{2}\times\left(\left[\frac{{\rm d}E}{M} \frac{{\rm d}\mu B_0}{M} {\rm d}P_\zeta{\rm d}\tau{\rm d}\xi{\rm d}\zeta\right]_{\sigma_{\rm HFS}}\right. \nonumber
\\
&\quad\, \left.+ \left[\frac{{\rm d}E}{M} \frac{{\rm d}\mu B_0}{M} {\rm d}P_\zeta{\rm d}\tau{\rm d}\xi{\rm d}\zeta\right]_{\sigma_{\rm LFS}}\right).
\label{eq:dvdx_can}
\end{align}

\noindent The index $\sigma = v_\parallel J_\parallel/|v_\parallel J_\parallel|$ determines in which domain (co- or counter-going) a volume element lies, irrespective of the orbit type (trapped or passing). Using the fact that canonical volume elements are identical anywhere on a GC orbit $(E,\mu,P_\zeta)$, Eq.~(\ref{eq:dvdx_can}) adds the volume elements at the high- and low-field side (HFS, LFS) midplane crossings and divides the result by 2, in accordance with our double-counting convention in Eq.~(\ref{eq:n_double}). Integrating over the angles $\xi$ and $\zeta$, using the pitch coordinate $\Lambda = \mu B_0/E$ instead of the magnetic moment, and applying our normalizations $\hat{v} = v/v_0$ and $\hat{E} = E/(Mv_0^2)$, we obtain
\begin{align}
&(2\pi)^2 \hat{E} {\rm d}\hat{E}{\rm d}\Lambda \times\left(\left[{\rm d}P_\zeta {\rm d}\tau\right]_{\sigma_{\rm HFS}} + \left[{\rm d}P_\zeta {\rm d}\tau\right]_{\sigma_{\rm LFS}}\right)\times\frac{v_0}{2} \nonumber
\\
&\approx (2\pi)^2 \hat{E}_i \Delta\hat{E}_i\Delta\Lambda_j \Delta P_{\zeta,k} v_0\Delta\tau_{ijkl} \times \frac{1}{2},
\label{eq:dcan}
\end{align}

\noindent for our CoM mesh with indices $(i,j,k)$ as defined in Section~\ref{sec:mesh}, where the cell index $k$ covers positive and negative signs of $\sigma$ for both the HFS and LFS midplane crossings. It remains to specify the increment $\Delta P_\zeta$.

Since our CoM mesh in Fig.~\protect\ref{fig:mesh} is defined in the midplane, the canonical toroidal momentum $P_\zeta$ for fixed $E$ and $\Lambda$ is a function of major radius only, which is here expressed in terms of $X = R - R_0$:
\begin{equation}
[P_\zeta]_{E\Lambda}(X) = -\frac{\Psi_{\rm P}(X)}{B_0} + \rho_0 \frac{\hat{v}_\parallel(X)}{B(X)} B_\zeta(X).
\label{eq:pzeta_x}
\end{equation}

\noindent Its increment can then be evaluated as
\begin{equation}
\Delta P_{\zeta,ijk} \approx P_\zeta\left(\left.X^{\rm (g)}_{k+1}\right|\hat{E}_i,\Lambda_j\right) - P_\zeta\left(\left.X^{(g)}_k\right|\hat{E}_i,\Lambda_j\right),
\label{eq:dp1}
\end{equation}

\noindent where $X^{(g)}_k$ and $X^{(g)}_{k+1}$ are the grid points adjacent to cell $X_k$. Alternatively, using $v_\parallel = \sigma_B\sqrt{2E(1 - \Lambda\hat{B})}$, the increment $\Delta P_\zeta$ can be expressed in terms of $\Delta X$ as
\begin{equation}
\Delta P_\zeta = \underbrace{\left|-\frac{\Psi_{\rm P}'}{B_0} - \frac{\rho_0}{\hat{v}_\parallel} \left(\hat{E} + \frac{\hat{v}_\parallel^2}{2}\right)\frac{B_\zeta B'}{B^2} + \rho_0 \frac{\hat{v}_\parallel}{B} B_\zeta'\right|}\limits_{|P_\zeta'|} \Delta X,
\label{eq:dp2}
\end{equation}

\noindent where radial derivatives like $P_\zeta' \equiv [\partial_X P_\zeta]_{E\Lambda}$ are taken along the midplane with fixed $E$ and $\Lambda$, and they are evaluated at cell centers $X_k$.

In summary, substituting Eq.~(\ref{eq:dt}) into (\ref{eq:dcan}) and multiplying by the number $N_\tau$ of markers used to represent an orbit, we obtain the GC phase space volume element represented by the orbit sample in cell $(E_i,\Lambda_j,X_k)$ in a form that can be readily evaluated numerically:\footnote{Eq.~(\ref{eq:vorb}) differs from the formula for the volume element given in Eq.~(40) of Ref.~\protect\cite{Bierwage12a}. We consider the present derivation to be more rigorous. This affects mainly trapped particle orbits, which were effectively absent in previous applications of the code {\tt VisualStart}.}
\begin{equation}
[\Delta\V_{\rm orb}]_{ijk} = 4\pi^2 v_0 \hat{E}_i \Delta\hat{E}_i\Delta \Lambda_j \underbrace{[\tau_{\rm pol}\Delta P_\zeta]_{ijk}}\limits_{\parbox{1.8cm}{\centering\scriptsize contains zeros \& singularities}}\times\frac{1}{2}.
\label{eq:vorb}
\end{equation}

\noindent The final factor $1/2$ originates from Eq.~(\ref{eq:n_double}) via Eq.~(\ref{eq:dvdx_can}) and compensates our double-counting of all orbits in the domain $X_{\rm min} \leq X \leq X_{\rm max}$ in Fig.~\ref{fig:mesh}(b) via index $k$.

\begin{figure}
\centering
\includegraphics[width=0.47\textwidth]{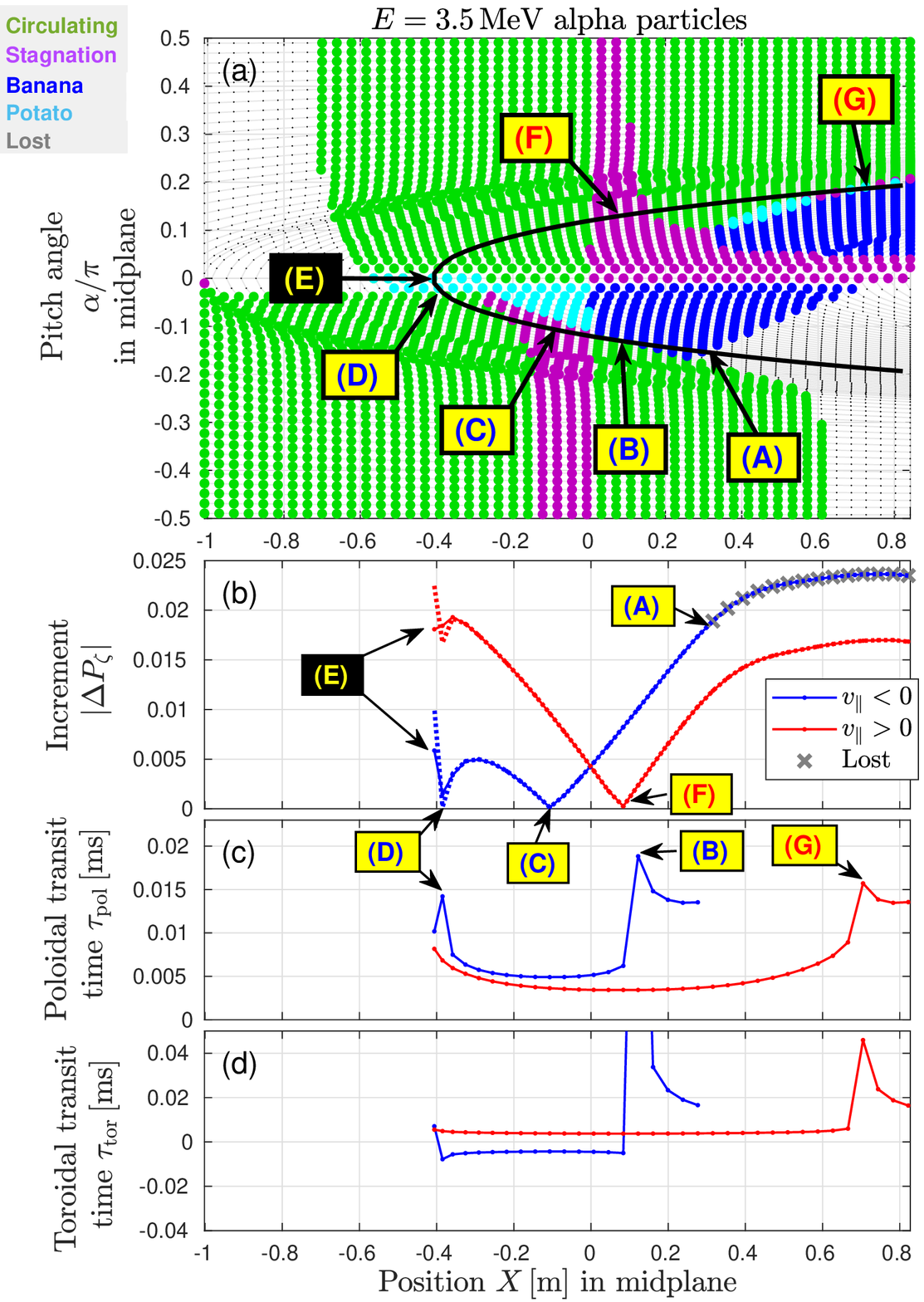}
\includegraphics[width=0.47\textwidth]{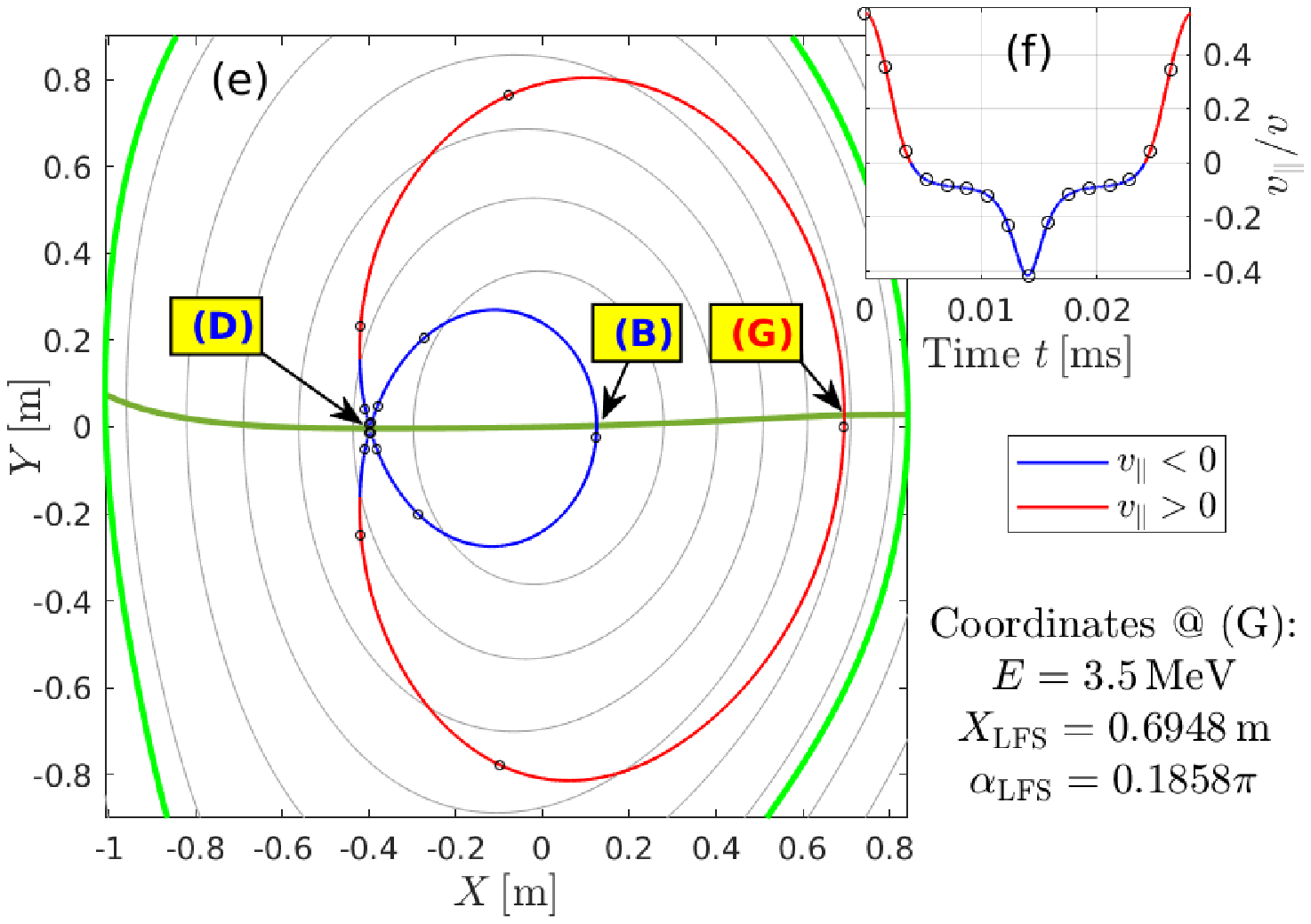}
\caption{Radial dependence of the increment $\Delta P_\zeta$ and the poloidal and toroidal transit times $\tau_{\rm pol}$ and $\tau_{\rm tor}$ on a contour $(E,\Lambda) = {\rm const}$. Panel (a) shows a CoM mesh constructed in the midplane using cell numbers $N_\Lambda = N_X = 48$ for alpha particles with $E = 3.5\,{\rm MeV}$. The total number of $\Lambda = {\rm const}$.\ contours is $(N_\Lambda/2) + N_X = 72$ and we inspect contour $j=40$ (counted from $\alpha = -\pi/2$), which is drawn as a black curve in (a). In panel (b), solid lines represent $|\Delta P_\zeta|$ given by Eq.~(\protect\ref{eq:dp1}) and the dotted lines are the alternative form $|P_\zeta'\Delta X|$ in Eq.~(\protect\ref{eq:dp2}). Blue and red lines in (b)--(d) represent orbits starting with $v_\parallel \lessgtr 0$, respectively. Labeled arrows: (A) loss boundary, (B,D,G) t-p boundaries, (C,D,F) stagnation points, and (E) the $v_\parallel = 0$ singularity. In fact, points (B,D,G) represent the same point in CoM space. This can be seen in panel (e), which shows the trajectory of a barely trapped orbit near the t-p separatrix. Panel (f) shows the time trace of $v_\parallel/v$ on this orbit, and $v_\parallel \lessgtr 0$ legs are colored blue and red.}
\label{fig:com_increment_40}%
\end{figure}

\subsection{Discussion}
\label{sec:weight_discuss}

The formula for the volume element in Eq.~(\ref{eq:vorb}) may look harmless, but the factor $\tau_{\rm pol} \Delta P_\zeta$ should be evaluated with care so as to minimize numerical inaccuracies. We have already seen in Fig.~\ref{fig:tp-boundary}(c,d) that $\tau_{\rm pol}$ possesses singularities at t-p boundaries, although not all of them are problematic as we will see shortly. The existence of zeros and singularities in the increment $\Delta P_\zeta$ can be readily anticipated from the derivative $|P_\zeta'|_{E\Lambda}$ in Eq.~(\ref{eq:dp2}), which can be approximated at leading order as
\begin{equation}
P_\zeta'(X) \approx -\frac{\Psi_{\rm P}'(X)}{B_0} + \frac{\rho_0}{\hat{v}_\parallel(X)} \left(\hat{E} + \frac{1}{2}\hat{v}_\parallel^2(X)\right).
\label{eq;dp2_x_approx}
\end{equation}

\noindent Clearly, $v_\parallel = 0$ constitutes a singularity and there are at least two points where the terms on the right-hand side cancel, giving $|P_\zeta'|_{E\Lambda} = 0$. Figure~\ref{fig:com_increment_40} shows an example for $3.5\,{\rm MeV}$ alpha particles. We consider orbits along the black parabolic line in Fig.~\ref{fig:com_increment_40}(a), where $\Lambda = 0.865$.

Figure~\ref{fig:com_increment_40}(b) shows the form of the increment $\Delta P_\zeta(X)$ evaluated using Eq.~(\ref{eq:dp1}) (solid curves) and Eq.~(\ref{eq:dp2}) (dotted). The values agree nicely nearly everywhere, except in the vicinity of the $v_\parallel = 0$ singularity labeled (E).\footnote{We expect that the red and blue curves in Fig.~\ref{fig:com_increment_40}(b,c) connect smoothly near $X \approx -0.4\,{\rm m}$ if one adds grid points closer to $v_\parallel = 0$.}
A verification exercise similar to that performed later in Section~\ref{sec:viz_map} showed systematic errors when using Eq.~(\ref{eq:dp2}). These errors seemed to be effectively absent when $\Delta P_\zeta$ is evaluated using Eq.~(\ref{eq:dp1}), which is therefore our default choice.

The zeros of $\Delta P_\zeta(X)$ correspond to stagnation points and imply that these and nearby orbits have a small weight factor $W_{ijk}$. In the example in Fig.~\ref{fig:com_increment_40} there are three such points: (C) and (F) are the O-type stagnation points in the domains of counter- and co-passing particles; whereas (D) is an X-type stagnation point situated on the separatrix of the t-p boundary.

The zero of $\Delta P_\zeta(X)$ at point (D) plays an important role: it cancels one of the three singularities of the poloidal transit time $\tau_{\rm pol}$ in Fig.~\ref{fig:com_increment_40}(c). In fact, a glance at the nearby barely trapped orbit in Fig.~\ref{fig:com_increment_40}(e) shows that points (B,D,G) actually refer to the same transit time singularity; i.e., the same t-p separatrix. Strictly speaking, our algorithm counts such separatrices thrice, but one of these instances --- namely the X-point at (D) --- vanishes, owing to $\Delta P_\zeta(X)$ being zero at that location. This confirms that all orbits are effectively double-counted, so that the overall factor $1/2$ appearing in Eq.~(\ref{eq:vorb}) is justified even near the t-p boundary. Of course, on a discrete mesh the cancellation of the third sample at point (D) is unlikely to be perfect, so some amount of numerical inaccuracy must be expected.

The remaining transit time singularities --- namely points (B) and (G) in our example --- have the consequence that orbits representing those and nearby portions of GC phase space may accumulate a relatively large weight. One could compensate these singularities by demanding that the value of the distribution function $f_{\rm orb}$ vanishes at the t-p boundary; e.g., by including a factor $\tau_{\rm pol}^{-1}$ in $f_{\rm orb}$. However, this may not be physically meaningful as the following arguments show.

The particles populating the barely trapped orbit in Fig.~\ref{fig:com_increment_40}(e) as well as nearby potato and counter-passing orbits spend a relatively large amount of time in a small portion of pitch-position space $(v_\parallel/v,R,z)$ near V-type stagnation points. This fact is highlighted by the bunching of the small black circle symbols in Fig.~\ref{fig:com_increment_40}(e,f), which represent 16 marker particles distributed uniformly in time. However, unlike the tiny O-type stagnation orbits around points (C) and (F), the large orbits near t-p boundaries like (B) and (G) can actually become densely populated in real plasmas. This becomes evident if one thinks in terms of particle sources: particles deposited on orbits located near a t-p boundary, like our example in Fig.~\ref{fig:com_increment_40}(e), will be conveyed to the V-type stagnation points and spend a long time in that region, producing a local spike in the particle density.

Of course, the finite particle supply rate along with instabilities and collisions will prevent the formation of a true singularity in a real plasma. Nevertheless, the above arguments show that, within certain limits, the spikes in the transit times in Fig.~\ref{fig:com_increment_40}(c) and resulting spikes in the orbit volume elements (\ref{eq:vorb}) and marker weights (\ref{eq:w_f_com}), are physical meaningful.

Such spikes can however cause numerical problems. In some cases, the resulting inaccuracies may be tolerable. In the worst case, the presence of singularities may prevent the attainment of numerical convergence, since a finer mesh will have grid points closer to the singularity. If necessary, one may adjust the weights as mentioned a few paragraphs earlier; e.g., by imposing an upper limit on their values (or their gradients) as in the flux limiter scheme of fluid dynamics. Although this trick may ensure convergence, it is likely to converge to a somewhat inaccurate value. On the other hand, one may argue that the correct value is unknown as it depends on physical mechanisms that have been ignored, such as scattering by collisions and field fluctuations. Perhaps the most elegant solution in the collisionless limit is to use an analytical estimate of a volume element near the singularity, provided that a well-behaved solution exists. Anyhow, we believe that the optimal approach to dealing with singularities depends on the application at hand, so we leave it for future research. It seems meaningful to deal with this problem in the context of smoothing techniques that will be needed to evaluate phase space gradients for instability analyses and may be based on physical scattering mechanisms.

\section{Modeling CoM distributions with large drifts}
\label{sec:mdl}

A computed or modeled equilibrium distribution $f_{\rm orb}$ is needed to determine the weight factor in Eq.~(\ref{eq:w_f_com}). A model that is a function of CoM only, such as $G_{\rm mdl}(\enr,\mu,P_\zeta)$, would directly yield an exact equilibrium distribution $f_{\rm orb}$. While simple models may suffice for physics studies at a fundamental level, predictive simulations and simulations used to interpret observations require more realism. It would be convenient to have $G_{\rm mdl}$ given in an explicit, purely analytical form. However, for fast ions with significant magnetic drifts, a large number of control parameters is usually necessary to mimic realistic distribution functions, which tends to make the modeling job cumbersome, at least for non-artificial intelligence.\footnote{With a human-manageable amount of control parameters, it can be difficult to design a CoM distribution that has both the desired radial profile and the desired pitch distribution, because both vary simultaneously when one varies parameters controlling the dependence of $f_{\rm mdl}$ on the canonical toroidal angular momentum $P_\zeta$.}

For instance, consider the problem of constructing a model for fusion-born alpha particles. Even if one assumes that their pitch angle distribution is uniform at birth time,\footnote{Precisely speaking, when an anisotropy exists in the distribution of fusion fuel particles, this anisotropy may also be imprinted on the fusion products due to angular correlations in nuclear reactions (e.g., see Ref.~\protect\cite{Deutsch51} and Fig.~10 in Ref.~\protect\cite{Hanson49}). Here this effect is neglected.\label{fn:alpha}}
strong pitch angle nonuniformities can promptly arise within one poloidal transit time. The loss cone, which contains orbits whose trajectories intersect plasma-facing components, is an obvious example. The magnetic drifts that cause those losses in toroidal geometry also affect the pitch angle distribution of confined orbits. For instance, for fusion-born alpha particles in ITER, the anisotropy associated with drift orbit topology was estimated to be around $10\%$ (see Section 7 of Ref.~\cite{Salewski18b}). Pitch anisotropies and bump-on-tail structures along the energy axis are also a well-known cause of velocity space instabilities that can be observed as ion cyclotron emissions (ICE) (e.g., see Refs.~\cite{Sumida19,Sumida21}).

\begin{figure}
[tb]
\centering
\includegraphics[width=0.47\textwidth]{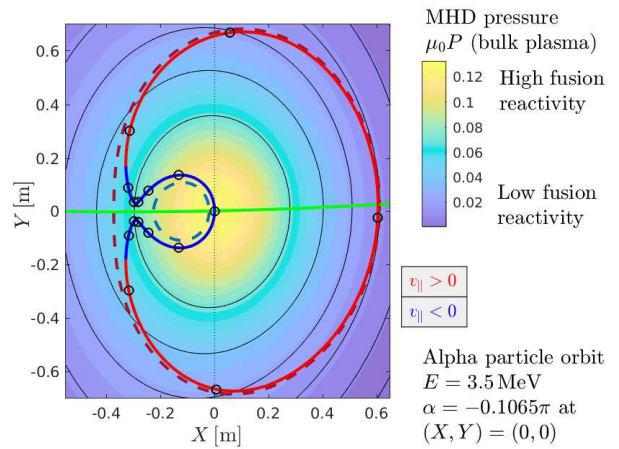}
\caption{GC orbit of a barely trapped $3.5\,{\rm MeV}$ alpha particle born at the magnetic axis $(X,Y) = (0,0)$ with pitch angle $\alpha = -0.1065\pi$. Small black circles indicate 16 marker particles separated by equal time intervals. At the nearby trapped-passing (t-p) boundary, this orbit decomposes into a small core-localized counter-passing orbit and a large potato orbit, which are indicated by dashed lines. Note that the outer orbit is disconnected from the hot central core, which appears yellow in the contours of the MHD pressure $P(\Psi_{\rm P})$ of the bulk plasma. This illustrates how the alpha particle distribution $f_{\rm orb}(\enr,\mu,P_\zeta)$ can develop a strong nonuniformity around the t-p boundary since the fusion reactivity is not a function of the alpha's $P_\zeta$ but of the bulk pressure (color contours) and, thus, magnetic flux $\Psi_{\rm P}$ (black contours).}
\label{fig:modelling}%
\end{figure}

Let us consider a more concrete example. Figure~\ref{fig:modelling} shows the GC orbit of a barely trapped $3.5\,{\rm MeV}$ alpha particle, which traverses both the hot plasma core and the cooler plasma periphery. If one uses a slowly varying function $G_{\rm mdl}(\enr,\mu,P_\zeta)$ to model the overall distribution of alpha particles in this range of energies and pitch angles, the density will be similar on the barely trapped orbit shown in Fig.~\ref{fig:modelling} and on the neighboring large potato and small counter-passing orbits indicated by dashed lines, because their coordinates have similar values.\footnote{In our example: $\mu B_0/E = 0.875...0.892$ and $\hat{P}_\zeta = 0.215...0.219$ with normalized $\hat{P}_\zeta \equiv \psi_{\rm P} - \hat{v}_\parallel \frac{B_\zeta}{B} \frac{\rho_0 B_0}{\Psi_{{\rm P},a} - \Psi_{{\rm P},0}}$ and $0 \leq \psi_{\rm P} \leq 1$.}
In other words, the particle density will be similar on both sides of the t-p boundary.

However, this may not be an accurate representation of a real distribution of fusion-born alphas, in particular when the fusion reactivity varies substantially in different portions of the orbit. To illustrate this, the $3.5\,{\rm MeV}$ alpha particle orbit in Fig.~\ref{fig:modelling} has been underlaid with the contours of the sharply peaked MHD pressure field from of our working example (cf.,~Fig.~\ref{fig:tok}), which is characteristic for the bulk component of a JET plasma where 3-ion RF heating was applied near the magnetic axis \cite{Nocente20, Kazakov21}. In such a case, fusion-born alphas from the plasma center will populate the entire orbit shown in Fig.~\ref{fig:modelling}, albeit in a nonuniform manner as indicated by the small black circles. Just across the t-p boundary, however, most of the newly born alphas will be concentrated on the small counter-passing orbit, while the large potato orbit would be populated only sparsely, since it is topologically disconnected from the particle source. In principle, this can produce a steep gradient in the distribution function $f_{\rm orb}(\enr,\mu,P_\zeta)$ across the t-p boundary. In reality, collisions or instabilities can be expected to have a smoothing effect, but depending on the relevant time scales, a strong nonuniformity may be sustained. Such information may be necessary for interpreting experimentally observed signals or for reproducing them in simulations, and it is advantageous to have the ability to construct suitable models.

A viable recipe for doing so can be found in Eq.~(\ref{eq:f_tavg}). Such an orbit time average can be readily realized using our orbit database and constitutes a physically meaningful way to convert an arbitrary model function $G_{\rm mdl}({\bm Z})$ into a true equilibrium distribution $f_{\rm orb}$:
\begin{align}
f_{\rm orb} =&\; \left<G_{\rm mdl}(\rho_{\rm P})\right>_{\rm orb} \equiv \frac{1}{\tau_{\rm orb}} \oint{\rm d}\tau\, G_{\rm mdl}({\bm Z}(\tau)) \nonumber
\\
\approx&\; f_{\rm orb}(i,j,k) = \frac{1}{N_{\tau,ijk}} \sum_{l=1}^{N_{\tau,ijk}} G_{\rm mdl}({\bm Z}_{{\rm gc},ijkl});
\label{eq:f_fmdl_avg}
\end{align}

\noindent where ${\bm Z}_{{\rm gc},ijkl}$ is the position of marker $l$ on orbit $(i,j,k)$. For instance, in the example discussed in the previous paragraphs, one may let $G_{\rm mdl} \propto P(\Psi_{\rm P})$, so that the MHD pressure field in Fig.~\ref{fig:modelling} acts as a particle source.

This approach can be viewed as taking the path of ``self-organization'', where we leave it to the equations of motion to determine a distribution function that is consistent with the magnetic geometry and system boundaries at hand, and the constraints encoded in $G_{\rm mdl}$ determine the particular spatial profile and velocity distribution for a certain case. This idea also underlies Monte-Carlo codes that follow GC orbits many times around the torus in the presence of sources and collisions. We mimic a part of this process with reduced computational effort: the GC orbits in our database are followed for only one poloidal turn in the absence of any perturbations (field fluctuations and collisions).

\begin{table*}[tb]
\begin{tabular}{c|cccc|c@{\hspace{0.24cm}}c|c@{\hspace{0.24cm}}c|c@{\hspace{0.24cm}}c}
\hline\hline
Form of & \multicolumn{4}{c|}{Sampling parameters} & \multicolumn{2}{c|}{$E_0 = 35\,{\rm keV}$} & \multicolumn{2}{c|}{$E_0 = 350\,{\rm keV}$} & \multicolumn{2}{c}{$E_0 = 3.5\,{\rm MeV}$}\\
$E$-distrib.\ & $N_E$ & $N_\alpha$ & $N_X$ & $N_\tau$ & $N_{\rm orb}$ & $N_{\rm mrk}$ & $N_{\rm orb}$ & $N_{\rm mrk}$ & $N_{\rm orb}$ & $N_{\rm mrk}$ \\
\hline
Mono-$E_0$ & $1$ & $2\times 64$ & $2\times 64$ & $128 \frac{L_{\rm orb}}{L_{\rm bnd}}$ (min.: 16) & 33,335 & 1.88M & 31,841 & 1.70M & 27,684 & 1.29M \\
Flat-$E \leq E_0$ & $24$ & $2\times 24$ & $2\times 24$ & $48 \frac{L_{\rm orb}}{L_{\rm bnd}}$ (min.: 16) & 112,483 & 2.74M & --- & --- & 99,308 & 2.18M \\
\hline\hline
\end{tabular}
\caption{Phase space sampling parameters in our five test cases. $N_E$, $N_\alpha$, $N_X$ determine the number of cells in $(E,\alpha,X)$-space defined in Section~\protect\ref{sec:mesh}. The number $N_\tau$ of markers sampling an orbit is (with a lower bound of 16) chosen to be proportional to the orbit length $L_{\rm orb}$, the reference value being the number of markers per plasma boundary length, here $L_{\rm bnd} \approx 7.50\,{\rm m}$. There are $N_{\rm mrk}$ markers sampling $N_{\rm orb}$ confined orbits.}
\label{tab:parm_mesh}
\end{table*}

\begin{table}[tb]
\begin{tabular}{c|c@{\hspace{0.27cm}}c@{\hspace{0.27cm}}c@{\hspace{0.27cm}}c@{\hspace{0.27cm}}c}
\hline\hline
$E\,[{\rm keV]}$ & Circ. & Stagn. & Banana & Potato & Lost $\approx$ \\
\hline
$35$ & 25,780 & 359 & 7,158 & 38 & 780 \\
$350$ & 24,430 & 1,066 & 6,191 & 154 & 2,280 \\
$3500$ & 20,531 & 3,165 & 3,505 & 483 & 6,430 \\
\hline
$0...35$ & 87,368 & 612 & 24,456 & 47 & 1,710 \\
$0...3500$ & 75,178 & 7,208 & 15,861 & 1,061 & 14,880 \\
\hline\hline
\end{tabular}
\caption{Partition of orbit types (circulating, stagnation, banana, potato, lost) as defined in Section~\protect\ref{sec:mesh}. These numbers are {\it not} weighted by particle densities, so they only characterize the partition of orbit samples on our mesh. The actual prompt losses of real particles are negligible in our large plasma with centrally peaked density profile.}
\label{tab:orb_type}
\end{table}

\section{Application and verification}
\label{sec:viz}

In this section, we demonstrate how the methods that we presented in this paper can be applied to a practical example. Section~\ref{sec:viz_mdl} describes the setup and how we construct --- from a simple model $G_{\rm mdl}$--- an equilibrium distribution $f_{\rm orb}$ of alpha particles that are assumed to be born near the center of a large tokamak plasma with an isotropic distribution of pitch angles at birth time. The results are discussed in Sections~\ref{sec:viz_monoE} and \ref{sec:viz_flatE}. In the final Section~\ref{sec:viz_map}, we demonstrate how we use our orbit database to convert a binned 4-D distribution function $f^{(0)}(E,\lambda,R,z)$ (or any other coordinates) back to an orbit-based representation $f_{\rm orb}^{(1)}$ in CoM space. The result is binned again to give $f^{(1)}$ for comparison with the original $f^{(0)}$. The procedure is then iterated one more time, yielding $f^{(2)}$, in order to reveal systematic errors. The contents of this section may be summarized in diagrammatic form like this:
\begin{equation}
\begin{array}{r@{\hspace{0.2cm}}c@{\hspace{0.cm}}c@{\hspace{0.2cm}}l}
\parbox{1cm}{\centering\text{\scriptsize Modeling}} & \parbox{2cm}{\centering\text{\scriptsize CoM space}} & & \parbox{2cm}{\centering\text{\scriptsize Arbitrary coordinates}} \\
\fbox{$G_{\rm mdl}$} \rightarrow & \fbox{$f_{\rm orb}^{(0)}(i,j,k)$} & \rightarrow & \fbox{$f^{(0)}(E,\lambda,R,z)$} \\
& & \swarrow & \;\; \vertapprox \;?\;\; \text{\scriptsize 1st verification} \\
& \fbox{$f_{\rm orb}^{(1)}(i,j,k)$} & \rightarrow & \fbox{$f^{(1)}(E,\lambda,R,z)$} \\
& & \swarrow & \;\; \vertapprox \;?\;\; \text{\scriptsize 2nd verification} \\
& \fbox{$f_{\rm orb}^{(2)}(i,j,k)$} & \rightarrow & \fbox{$f^{(2)}(E,\lambda,R,z)$}
\end{array}
\label{eq:map}
\end{equation}

\noindent Strictly speaking, we transform only between 2-D and 3-D spaces, $(A_j,P_\zeta(X_k)) \leftrightarrow (\lambda,R,z)$, since both sets share the energy coordinate $E$, which is a conserved quantity, so the mapping in that direction is identical. We will work here with an arbitrary normalization (henceforth indicated by hats), so the values of $\hat{f}_{\rm orb}$ of $\hat{f}$ must match but bear no particular meaning.

Our distribution functions will be inspected here in the same way as an experimental diagnostician would: measure spatial distributions by integrating over certain portions of velocity space, and measure velocity distributions by integrating over certain portions of position space. The required binning operations and Jacobians are described in \ref{apdx:bin}.

\begin{figure}
[tb]
\centering
\includegraphics[width=0.47\textwidth]{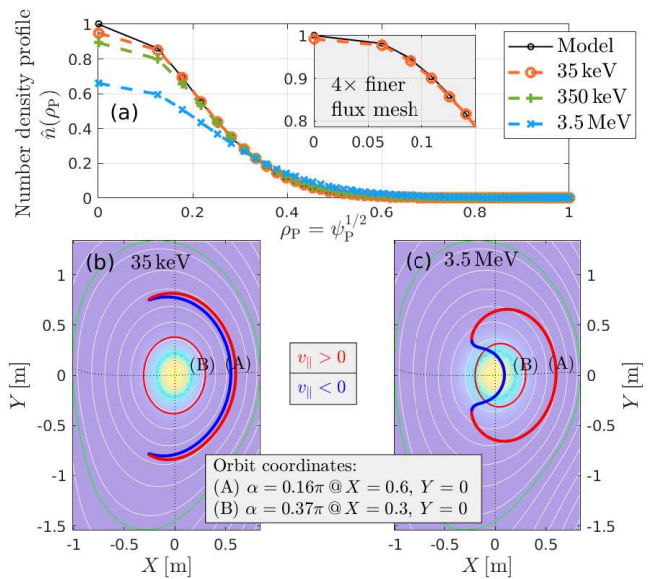}
\caption{Panel (a) shows the radial profiles of our source model $\hat{n}_{\rm mdl}(\rho_{\rm P})$ used in Eq.~(\protect\ref{eq:f_mdl}) (black) and constructed mono-energetic alpha particle densities $\hat{n}(\rho_{\rm P})$ (colored) as functions of the square root of normalized poloidal flux, $\rho_{\rm P} = \psi_{\rm P}^{1/2}$, on a grid that is uniform in $\psi_{\rm P}$. For $35\,{\rm keV}$ and $3.5\,{\rm MeV}$, panels (b) and (c) show the corresponding density fields $\hat{n}(X,Y)$ in the poloidal plane, overlaid with examples of (A) trapped and (B) passing orbits as well as flux surfaces (gray).}
\label{fig:viz_profs_orbits}%
\end{figure}

\subsection{Setup and parameters}
\label{sec:viz_mdl}

In order to highlight how toroidal geometry affects the form of the constructed equilibrium distribution function via the mirror force and magnetic drifts, we start from a quasi-distribution $G_{\rm mdl}$ that is independent of pitch and has a relatively sharp peak in minor radius:
\begin{equation}
G_{\rm mdl}(\rho_{\rm P}) = C\times \hat{n}_{\rm mdl}(\rho_{\rm P}) \times H_{\rm mdl}(E),
\label{eq:f_mdl}
\end{equation}

\noindent where $C$ is a normalization constant that is arbitrary here.

The radial density profile $\hat{n}_{\rm mdl}(\rho_{\rm P})$ is shown as a solid black line in Fig.~\ref{fig:viz_profs_orbits}(a). Although this is primarily meant to be a toy model for testing our algorithms with a localized particle source and large magnetic drifts, we note that this profile is physically feasible if one considers scenarios with a sharply peaked ion temperature profile as produced in 3-ion-heated JET plasmas \cite{Nocente20,Kazakov21}.

The energy dependence of particle distributions in a real plasma is governed by processes such as collisional drag and instabilities (e.g., see Fig.~4 in Ref.~\cite{Bierwage18}). These processes are not considered here. We use energy only to exemplify the effects of magnetic drifts, which increase with $E$; or, rather, with its square root $\sqrt{E}$, because the {\it effect} of magnetic drifts is roughly proportional to the ratio $v_{\rm dB}/v$ of the drift velocity to the particle velocity. For this purpose, we use two types of distributions: one mono-energetic with $E = E_0$, and one that is flat in the range $0 \leq E \leq E_0$ and zero beyond:
\begin{align}
H_{\rm mdl}(E) &= \left\{
\begin{array}{l}
\delta(E - E_0) \quad\;\;\; = \text{ mono-energetic}, \\
\frac{1 - \int_0^E{\rm d}E'\,\delta(E' - E_0)}{E_0} = \frac{1}{E_0}\,{\rm for}\, 0 \leq E \leq E_0,
\end{array}
\right.
\label{eq:e_mdl}
\end{align}\vspace{-0.3cm}

\noindent where $\delta$ is the Dirac delta distribution.

Table~\ref{tab:parm_mesh} shows the parameters for five test cases that we will consider here: three mono-energetic alpha distributions with $E = E_0 = (35,350,3500)\,{\rm keV}$ and two distributions that are uniform in the energy range $0 \leq E \leq E_0$ with $E_0 = (35,3500)\,{\rm keV}$. Table~\ref{tab:orb_type} summarizes the numbers of circulating, stagnation, banana, potato and lost orbit samples (CoM mesh points) in each case. Panels (b) and (c) of Fig.~\ref{fig:viz_profs_orbits} show how the contours of a banana orbit (A) and a circulating orbit (B) change due to magnetic drifts when the kinetic energy is increased from $35\,{\rm keV}$ to $3.5\,{\rm MeV}$.

While the model in Eq.~(\ref{eq:f_mdl}) has a simple appearance, the actual form of the alpha particle distributions we construct here is determined by the combination of Eq.~(\ref{eq:f_mdl}) and the equations of GC motion (\ref{eq:dxgc_dt}) and (\ref{eq:du_dt}), whose solutions are folded into the simple model via the integral (\ref{eq:f_fmdl_avg}). In other words, the geometric effects are captured by the GC orbits in our database, whereas the model $G_{\rm mdl}$ effectively describes the structure of the particle sources that populate those orbits, and the integral (\ref{eq:f_fmdl_avg}) combines all that information.

The geometric effects that we expect to see in this setup may be summarized as follows. These properties of a toroidally confined and radially bounded plasma are well-known and are listed here only for completeness and to simplify the discussion of the results:
\begin{itemize}
\item  The peripheral plasma will be populated by particles on orbits that also pass through the core as in Fig.~\ref{fig:viz_profs_orbits}(c). Such orbits are found in a certain range of pitch angles around the t-p boundary.
\item  Flux surface-averaged profiles tend to be smoothed on the scale length of the magnetic drifts. The gyroradii have a similar effect and may be included via satellite particles \cite{Bierwage16c} (not done here).
\item  Magnetic drifts shift co-passing orbits outward in $R$ and counter-passing orbits inward.
\item  The mirror force confines some particles to the low-field side of the plasma and causes them to accumulate near V-type stagnation points.
\item  The energy- and pitch-dependence of the above-mentioned spatial nonuniformities gives rise to net flows both poloidally (due to radial nonuniformity) and toroidally (due to a local imbalance between co- and counter-going particles).
\end{itemize}

\subsection{Properties of mono-energetic distributions}
\label{sec:viz_monoE}

Figure~\ref{fig:viz_profs_orbits}(a) shows the computed density profiles $\hat{n}(\rho_{\rm P})$ on a mesh that is uniformly spaced in normalized poloidal flux $\psi_{\rm P} = \rho_{\rm P}^2$. The profile for $35\,{\rm keV}$ alphas is similar to the model $\hat{n}_{\rm mdl}(\rho_{\rm P})$. The small discrepancy at $\rho_{\rm P} = 0$ can be eliminated with a finer mesh as shown in the inset. The magnetic drifts increase with increasing energy, causing the profile $\hat{n}(\rho_{\rm P})$ to broaden and the central value $\hat{n}(0)$ to drop. The total number of particles in the plasma is nearly the same in all three cases, since prompt losses are relatively small: less than $1\%$ for $350\,{\rm keV}$ and about $2.5\%$ for $3.5\,{\rm MeV}$.

\begin{figure}
[tb]
\centering
\includegraphics[width=0.465\textwidth]{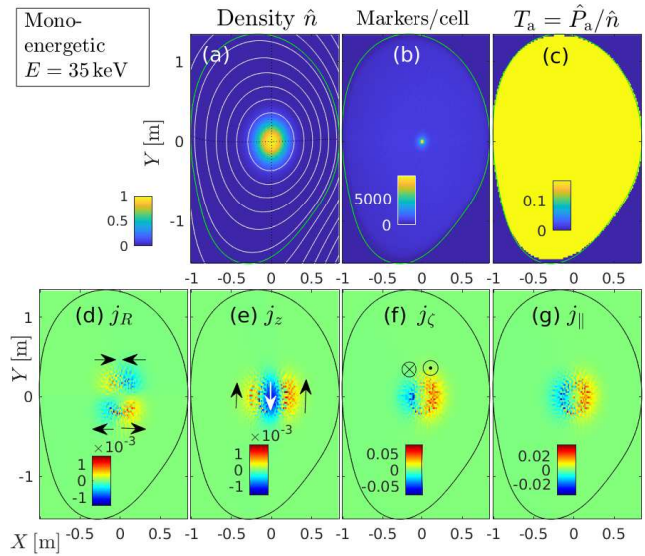}
\caption{Moments of the mono-energetic alpha distribution with $35\,{\rm keV}$. We integrated over velocity space using Eq.~(\ref{eq:f_rz}) with the following weight factors: (a) $w$ for particle density $\hat{n}(X,Y)$; (b) $1$ for the number of markers per cell; (c) $w\times E/3$ for alpha particle pressure $\hat{P}_{\rm a}$, then divided by density to yield temperature $T_{\rm a} = \hat{P}_{\rm a}/\hat{n}$; (d)--(g) $w\times (v_R,v_z,v_\zeta,v_\parallel)$ for horizontal, vertical, toroidal and parallel current densities $(j_R,j_z,j_\zeta,j_\parallel)$. The amplitude is normalized to $\hat{n}_0 = 1$.}
\label{fig:viz_35k_x-space}%
\end{figure}

\begin{figure}
\centering
\includegraphics[width=0.465\textwidth]{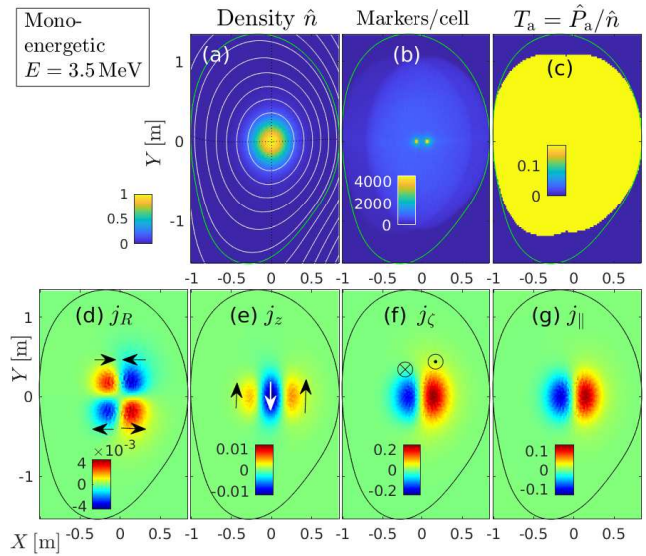}
\caption{Moments of the mono-energetic alpha distribution with $3.5\,{\rm MeV}$. Arranged as Fig.~\protect\ref{fig:viz_35k_x-space}.}
\label{fig:viz_3500k_x-space}%
\end{figure}

Figures~\ref{fig:viz_35k_x-space} and \ref{fig:viz_3500k_x-space} show the form of various moments of the alpha distributions with kinetic energies $35\,{\rm keV}$ and $3.5\,{\rm MeV}$ in the poloidal plane $(X,Y)$:
\begin{itemize}
\item  Particle density (a): Even for $3.5\,{\rm MeV}$, our model yields an alpha density that is peaked at the axis, but the peak is wider than in the $35\,{\rm keV}$ case, especially in the horizontal ($X = R - R_0$) direction.

\item  Markers/cell (b): Since we have set a lower bound of 16 markers per orbit, the smallest stagnation orbits stand out in the plots showing the number of markers per cell, while the surroundings are fairly uniform. At $35\,{\rm keV}$, the co- and counter-passing stagnation points effectively coincide with the magnetic axis. At $3.5\,{\rm MeV}$, their mean separation along $X$ is about $0.14\,{\rm m}$ (cf.~Fig.~\ref{fig:com_increment_40}(b)).

\item Temperature (c): With the density nonuniformity divided out, the $3.5\,{\rm MeV}$ temperature field clearly shows that our alpha particle distribution effectively consists of two partially overlapping co- and counter-going mono-energetic beams, which are shifted radially out- and inward, respectively.

\item Flows (d--g): The relative shift of the co- and counter-going beams causes net local currents. The poloidal current field (consisting of $j_R$ and $j_z$) forms two vortices, one on the high-field side (HFS) of the magnetic axis rotating clockwise, and one on the low-field side (LFS) rotating counter-clockwise. Toroidal and parallel currents $j_\zeta$ and $j_\parallel$ are very similar here, since the magnetic pitch is relatively uniform ($q \approx 1$) in the plasma center. The structure and spatial extent of the currents is similar at lower and higher energies, but the magnitude should be proportional to $\sqrt{E}$ and, thus, differ by about a factor of 10. This seems to be the case here if one accounts for the fact that the $35\,{\rm keV}$ case has a smaller signal-to-noise ratio.
\end{itemize}

\begin{figure}
\centering
\includegraphics[width=0.47\textwidth]{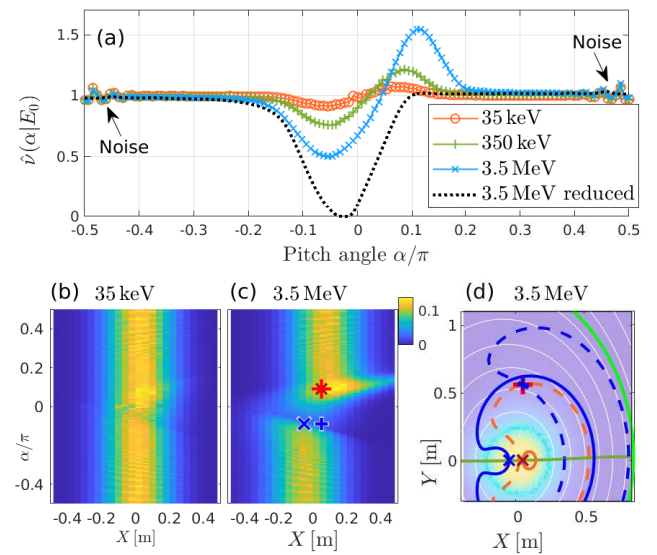}
\caption{Pitch angle distributions for the mono-energetic cases. Panel (a) shows pitch profiles $\hat{\nu}(\alpha|E_0)$ as defined in Eq.~(\protect\ref{eq:f_v}). The black dotted curve shows the $3.5\,{\rm MeV}$ result for a reduced domain size, where the artificial plasma boundary has been placed at $0.25\%$ of the flux space \protect\cite{Bierwage22b}, removing many large orbits (scaled to match the blue curve outside the loss cone). Panels (b) and (c) show contours of the local pitch distribution $\hat{\nu}(X,\alpha|E_0)$ obtained by integrating over the height of the spatial domain in radial bins of size $\Delta X = 0.04\,{\rm m}$. The two crosses and two plus symbols (sometimes overlapping) in panels (c) and (d) mark the starting points of four alpha particle orbits with $E_0 = 3.5\,{\rm MeV}$ whose poloidal contours are shown in panel (d). Blue and orange orbits start with $\alpha = -0.09\pi$ and $+0.09\pi$, respectively. Solid orbits start from $Y=0$ (crosses) and dashed orbits from $Y = 0.56\,{\rm m}$ (pluses). Panel (d) also shows the color contours of the $3.5\,{\rm MeV}$ alpha particle density field from Fig.~\protect\ref{fig:viz_3500k_x-space}(a), $\Psi_{\rm P}$ contours (white), the midplane (dark green), and the boundary (light green).}
\label{fig:viz_falpha_mono-E}%
\end{figure}

Figure~\ref{fig:viz_falpha_mono-E} shows the pitch angle distribution of our mono-energetic alpha particles. The spatially integrated distributions $\hat{\nu}(\alpha|E_0)$ in panel (a) show that a pitch nonuniformity exists around $\alpha = 0$, up to intermediate pitch angles, $|\alpha| \lesssim 0.2\pi$. The deviation from the mean increases from about $10\%$ for $35\,{\rm keV}$ to about $50\%$ for $3.5\,{\rm MeV}$. At larger pitch angles, $|\alpha| \gtrsim 0.2\pi$, the distributions are flat (except for aliasing noise, to be discussed later). More detailed views of the nonuniformities in the region $|\alpha| \lesssim 0.2\pi$ of the cases with $E_0 = 35\,{\rm keV}$ and $3.5\,{\rm MeV}$ are provided in panels (b) and (c), where we plot the contours of $\hat{\nu}(X,\alpha|E_0)$ to show the radial dependence of the pitch angle distribution in the range $-0.48\,{\rm m} < X < 0.48\,{\rm m}$. Here, the summation over markers was performed over the $Y$ coordinate for radial bins of size $\Delta X = 0.04\,{\rm m}$.

The structures seen in Fig.~\ref{fig:viz_falpha_mono-E} can be readily explained on the basis of our model function $G_{\rm mdl}$, the effects of toroidal geometry (mirror force and magnetic drifts), and the loss boundary. Rather than saying anything new, the purpose of the following discussion is to show that our procedures work reliably and produce the expected results. We focus on the $3.5\,{\rm MeV}$ case.

Figure~\ref{fig:viz_falpha_mono-E}(c) clearly shows the opposite radial shift of co- and counter-going particles due to their opposite magnetic drifts: the peak in the region $\alpha > 0$ is shifted towards the right ($X > 0$) and the peak in the region $\alpha < 0$ is shifted to the left ($X < 0$). The wedge-shaped structure of the pitch anisotropy around $\alpha = 0$ represents the widening of the domain of trapped particles towards the low-field side as expected from Fig.~\ref{fig:com_increment_40}(a).

In order to explain the structure of the pitch anisotropy in the region $|\alpha| \lesssim 0.2\pi$, it is helpful to consider the trajectories of a few GC orbits in that region. For this purpose, Fig.~\ref{fig:viz_falpha_mono-E}(d) shows the GC orbits of four simulation particles with energy $E = 3.5\,{\rm MeV}$. Two of them (solid lines) are launched near the midplane ($Y_{\rm start} \approx 0$): the small orange orbit starts from $(X,\alpha)_{\rm start} = (0.05\,{\rm m}, 0.09\pi)$, and the blue orbit starts from $(X,\alpha)_{\rm start} = (-0.05\,{\rm m}, -0.09\pi)$, as indicated by the crosses in panels (c) and (d). The blue orbit is a large potato orbit that passes both near the center of the plasma and through the periphery. The small orange orbit is a stagnation orbit. Now recall Eq.~(\ref{eq:f_fmdl_avg}), where we defined the orbit distribution $f_{\rm orb}$ to be the orbit time-average of the model $G_{\rm mdl}(\rho_{\rm P})$ that represents our alpha particle birth profile. Since the latter is sharply peaked near the plasma center, the values of $G_{\rm mdl}(\rho_{\rm P})$ can differ substantially on the above-mentioned pair of orbits.

Generally speaking, orbits that spend most of their time inside the peak of the source profile $G_{\rm mdl}(\rho_{\rm P})$ are densely populated with physical particles, so our GC markers representing them have larger weights $w_l$. Orbits that spend most of the time outside are sparsely populated. The mirror force and magnetic drifts have the consequence that particles on counter-passing orbits spend less time inside the central peak of $G_{\rm mdl}(\rho_{\rm P})$ than co-passing orbits. In addition, the losses of co- and counter-going particles are asymmetric, especially in the vicinity of the trapped-passing boundary. These effects are manifested in the minima and maxima of $\hat{\nu}(\alpha|E_0)$ and $\hat{\nu}(X,\alpha|E_0)$ in Fig.~\ref{fig:viz_falpha_mono-E}, which are most pronounced in the high-energy case due to its larger drifts.

Note that the distribution $\hat{\nu}(X,\alpha|E_0)$ in Fig.~\ref{fig:viz_falpha_mono-E}(c) was computed by integrating along the vertical coordinate $Y$, so it also includes particles on orbits like those shown by dashed lines in panel (d). The dashed orbits were launched from $(X,\alpha)_{\rm start} = (0.05\,{\rm m},\pm 0.09\pi)$, but more than half a meter above the midplane, from $Y_{\rm start} = 0.56\,{\rm m}$, as indicated by pluses in panels (c) and (d). The orbit starting with a positive pitch (orange) passes through the central plasma, where it acquires a relatively large weight due to the centrally peaked source profile $G_{\rm mdl}(\rho_{\rm P})$. In contrast, the orbit starting with a negative pitch travels outward and nearly hits the boundary (light green curve).

In fact, the entire peak of $\hat{\nu}(\alpha|E_0)$ around $\alpha/\pi \approx +0.1$ can be explained in that way. We have verified this with a test case where the plasma boundary had been placed at $\psi_{\rm P} = 0.25$ ($\rho_{\rm P} = 0.5$) instead of $\psi_{\rm P} = 1$, so that orbits are constrained to the inner 25\% of the present flux space \cite{Bierwage22b}. In that case, the plasma boundary is located around $X \approx 0.38\,{\rm m}$ at the height of the outer midplane, so that the three large orbits in Fig.~\ref{fig:viz_falpha_mono-E}(d) are all lost. The resulting pitch angle profile is shown as a black dotted line in Fig.~\ref{fig:viz_falpha_mono-E}(a). It has a deeper and wider loss cone, and there is no peak at $\alpha/\pi \approx +0.1$.

\begin{figure}
[tb]
\centering
\includegraphics[width=0.47\textwidth]{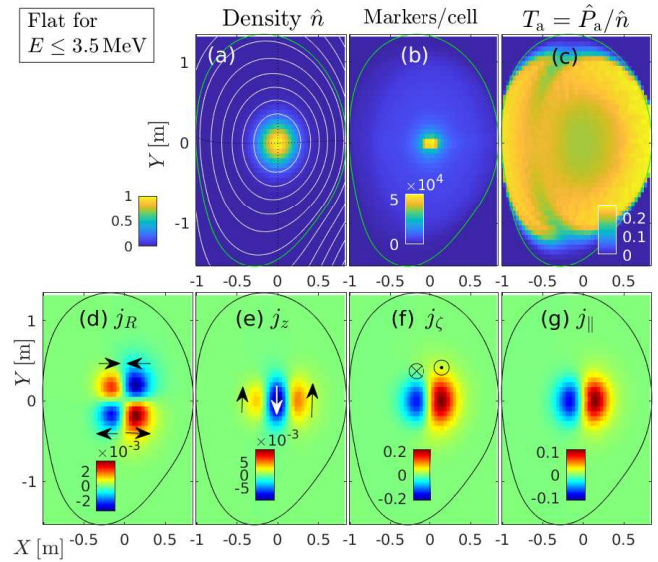}
\caption{Moments of the alpha distribution that is uniform in the energy range $E = 0...3.5\,{\rm MeV}$. Arranged as Fig.~\protect\ref{fig:viz_35k_x-space}.}
\label{fig:viz_0-3500k_x-space}%
\end{figure}

\begin{figure}
[tb]
\centering
\includegraphics[width=0.47\textwidth]{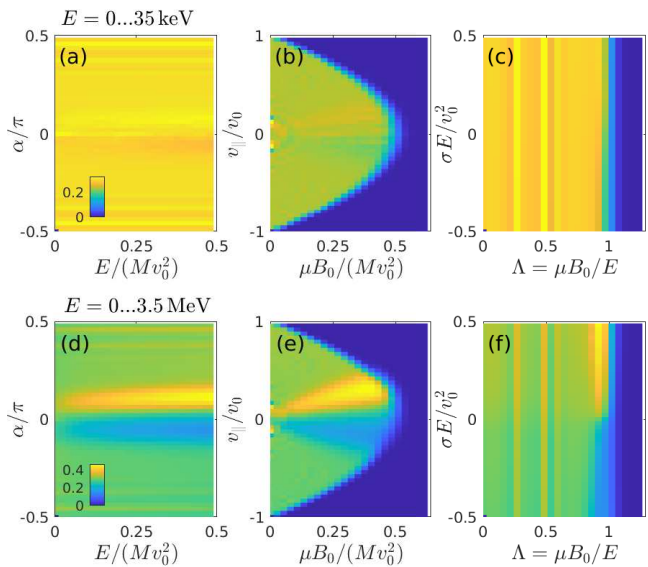}
\caption{Spatially integrated velocity distributions $\hat{\nu}(a,b) = \J^{-1}_{ab} \times \hat{h}_{\rm p}(a,b)$ in the cases that are uniform in the energy ranges $E = 0...35\,{\rm keV}$ (top) and $0...3.5\,{\rm MeV}$ (bottom), visualized in different coordinates: $\hat{\nu}(E,\alpha)$ on the left, $\hat{\nu}(\mu,v_\parallel)$ in the central column, and $\hat{\nu}(\Lambda,\enr)$ with $\enr \equiv \sigma E$ on the right. For plots of $\hat{\nu}(E,\lambda)$, see Fig.~\protect\ref{fig:viz_0-3500k_0th-1st-order}. Note that different color scales are used in the upper and lower row.}
\label{fig:viz_0-E0_v-space}%
\end{figure}

\subsection{Properties of flat-energy distributions}
\label{sec:viz_flatE}

Realistic alpha particle distributions are nonuniform in energy. The reason for our use of a model that is flat in energy is that the $E$-dependence of geometrically induced nonuniformities can be easily seen, and it is also easier to verify that our method works as expected.

The densities and current fields are very similar to those in the mono-energetic cases shown in Figs.~\ref{fig:viz_35k_x-space} and \ref{fig:viz_3500k_x-space} above. A few different features can be observed in the high energy case, $E = 0...3.5\,{\rm MeV}$, whose moments are shown in Fig.~\ref{fig:viz_0-3500k_x-space}. The presence of particles with a wide range of energies can be inferred primarily from the blurring of the stagnation points in panel (b), and from the nonuniformity of the temperature field in (c). The latter resembles somewhat the letters ``IO'' as fast particles dominate outside the central density peak; especially the co-passing ones (forming the ``O'' shifted to the right) for the same reasons as discussed in the last paragraphs of Section~\ref{sec:viz_monoE} above.

Figure~\ref{fig:viz_0-E0_v-space} shows the velocity distributions integrated over position space. One can see that, apart from noise, our result has the desired uniformity in both energy and pitch in the domains populated by passing particles. As discussed in Section~\ref{sec:viz_monoE} above, the pitch anisotropy around $\alpha = 0$ is induced by toroidal geometry (mirror force, magnetic drifts and resulting losses) in combination with the centrally peaked source profile. The anisotropy clearly increases with increasing energy.

Our orbit database is relatively sparse ($N_E\times N_\alpha \times N_X = 24\times 48\times 48$, see Table~\ref{tab:parm_mesh}), and the bins used in Figs.~\ref{fig:viz_0-3500k_x-space} and \ref{fig:viz_0-E0_v-space} have effectively the same resolution: $N_E\times N_\alpha \times N_{Rz} = 24\times 48\times 48^2$. The binned results are smooth along energy, because the orbit samples coincide with the bins. Most of the noise in Fig.~\ref{fig:viz_0-E0_v-space} is in the pitches $\alpha$ and $\Lambda$, where the distribution of bins is similar but not identical to the distribution of orbit samples (CoM space). In other words, the noise in the range $0.3 \lesssim |\alpha|/\pi \lesssim 0.5$ and $\Lambda \lesssim 0.7$ is effectively an {\it aliasing} effect. The noise is largely suppressed in the poloidal plane $(X,Y)$ in Fig.~\ref{fig:viz_0-3500k_x-space}, and in $(\hat{v}_\parallel,\hat{\mu})$-space shown in the central column of Fig.~\ref{fig:viz_0-E0_v-space}, since the bins in these coordinates have a shape very different from the mesh we used to sample the CoM space (cf.\ Figs.~\ref{fig:mesh} and \ref{fig:com_increment_40}).

Figure~\ref{fig:viz_0-E0_v-space} also highlights potential difficulties in attempts to convert the binned distributions from one set of coordinates to another. In particular, the small region around the origin $(\mu,v_\parallel) = (0,0)$ in the central column of Fig.~\ref{fig:viz_0-E0_v-space} is expanded to the full length of an axis in $(E,\alpha)$ and $(\Lambda,\enr)$ spaces on the left and right sides of Fig.~\ref{fig:viz_0-E0_v-space}, so some features of the distribution functions are likely to get lost in direct conversions unless a polar mesh is used in $(\mu,v_\parallel)$ coordinates, which is equivalent to using another set of coordinates, such as $(E,\alpha)$. We emphasize that we have {\it not} performed direct conversions between binned distributions in Fig.~\ref{fig:viz_0-E0_v-space}. They were all obtained by binning, on the respective mesh, the marker particles of the orbit-based representation $f_{\rm orb}$.

\begin{figure}
[tb]
\centering
\includegraphics[width=0.47\textwidth]{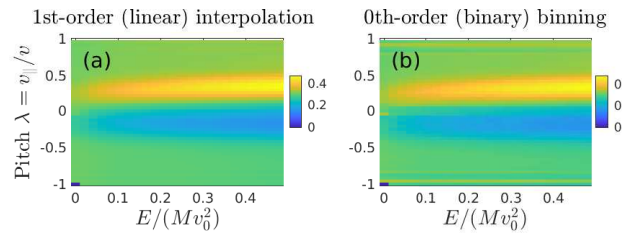}
\caption{Comparison between 1st- and 0th-order binning results $\hat{\nu}(E,\lambda)$ in the flat-energy case with $E = 0...3.5\,{\rm MeV}$. Here, we loaded $N_\tau = 480\times L_{\rm orb}/L_{\rm bnd}$ (min.\ 16) markers per orbit, giving a total of $N_{\rm mrk} \approx 18{\rm M}$ markers. Panel (b) is a projection of the input distribution $\hat{f}^{(0)}(E,\lambda,R,z)$ used in our verification exercise in Section~\protect\ref{sec:viz_map}.}
\label{fig:viz_0-3500k_0th-1st-order}%
\end{figure}

The binning in Figs.~\ref{fig:viz_profs_orbits}--\ref{fig:viz_0-E0_v-space} was performed using a 1st-order (linear) interpolation scheme as commonly used in particle-in-cell (PIC) codes, where each marker is assumed to have the shape of a top-hat function whose width is equal to the bin size. Naturally, the noise is larger with a 0th-order (binary) binning scheme, where markers resemble $\delta$ functions. The difference between 1st- and 0th-order binning can be seen by comparing panels (a) and (b) of Fig.~\ref{fig:viz_0-3500k_0th-1st-order}, where we show the spatially integrated velocity distribution $\hat{\nu}(E,\lambda)$ with $\lambda = v_\parallel/v = \sin^{-1}(\alpha)$. Note also that the result in Fig.~\ref{fig:viz_0-3500k_0th-1st-order}(a) appears smoother than that in \ref{fig:viz_0-E0_v-space}(d) simply by using $\lambda$ instead of $\alpha$, because this change in coordinates reduces the above-mentioned aliasing effect. Another difference is that Fig.~\ref{fig:viz_0-3500k_0th-1st-order} was obtained with 10 times more markers per orbit, but this does not have a strong influence on the aliasing in the domain of deeply passing particles, since those markers rarely cross the bin boundaries.

\subsection{Verification of reversibility and accuracy of coordinate conversions via an orbit-based representation}
\label{sec:viz_map}

The results discussed in the previous sections were obtained by binning the marker weights of our orbit-based representation in CoM space to a mesh in arbitrary coordinates. The reverse transformation is also easily performed as we will demonstrate in this section. On platforms like ITER IMAS, we expect this to be an essential part of the transformation toolbox, especially if an orbit-based representation is used as a reference. Here, this operation serves us primarily as a verification exercise to test with what degree of accuracy we can perform and reverse coordinate conversions without applying any smoothing algorithms.

We consider the flat-energy case with $E_0 = 3.5\,{\rm MeV}$ and follow the procedure that was outlined in Eq.~(\ref{eq:map}). Let us now define the relevant steps in detail:
\begin{enumerate}
\item  Using Eqs.~(\ref{eq:f_fmdl_avg}) and (\ref{eq:f_mdl}), we construct a distribution function $f_{\rm orb}$ in the orbit-based representation:
\begin{equation}
G_{\rm mdl} \rightarrow f_{\rm orb}(i,j,k).
\end{equation}

The CoM mesh that underlies our orbit database has the same number of cells as specified in Table~\ref{tab:parm_mesh}: $N_E\times (N_\alpha\times N_X) = 24\times48^2$. The number of markers per boundary length is increased by a factor 10 to $N_\tau = 480\times L_{\rm orb}/L_{\rm bnd}$ (min.\ 16), giving a total of $N_{\rm mrk} = 18\,{\rm M}$ markers.

\item The orbit-based CoM distribution $f_{\rm orb}$ is binned on a uniform 4-D mesh in the widely used coordinates $(E,\lambda,R,z)$. In compact form, this operation can be written as
\begin{equation}
f_{\rm orb}^{(\eta)}(i,j,k) \rightarrow f^{(\eta)}(E,\lambda,R,z)
\end{equation}
where $\eta = 0,1,2$ is the iteration number. The initial result $f^{(0)}$ is referred to as the ``input'' distribution for the subsequent back-and-forth transformations. More precisely, the binning operation we perform is given by Eq.~(\ref{eq:f_map}) in the form
\begin{align}
&f(E_i,\lambda_j,R_k,z_m) = \nonumber \
\\
&\sum_{n=1}^{\rm N_{\rm mrk}} \frac{w_n \times \Pi({\bm Z}_{{\rm gc},n} - {\bm Z}_{ijkm})}{(2\pi)^2 \left[\frac{B^*_\parallel}{B} \hat{v} R\right]_n \Delta E_i\Delta\lambda_j\Delta R_k\Delta z_m}.
\end{align}
where the shape factor $0 \leq \Pi \leq 1$ is a top-hat function, which means that the markers resemble $\delta$ functions. That is, we use the 0th-order binning scheme as in Fig.~\ref{fig:viz_0-3500k_0th-1st-order}(b) and the number of bins (= cells) is $N_E \times N_\lambda \times (N_R \times N_z) = 24\times 24\times 48^2$. The pitch resolution was reduced from $N_\lambda = 48$ to $24$ in order to reduce noise and aliasing (cf.~Section~\ref{sec:viz_flatE}).

\item  For each orbit in the database, we compute new weight factors $W_{ijk}^{(\eta+1)}$ by integrating $f^{(\eta)}(E,\lambda,R,z)$ in time $\tau$ along the orbit contour as
\begin{equation}
W_{ijk}^{(\eta+1)} = \left[\Delta\V_{\rm orb}\sum_{l=1}^{N_\tau} f^{(\eta)}(E,\lambda(\tau_l),R(\tau_l),z(\tau_l))\right]_{ijk},
\end{equation}
which gives a new orbit-based distribution:
\begin{equation}
f^{(\eta)}(E,\lambda,R,z) \rightarrow f_{\rm orb}^{(\eta+1)}(i,j,k).
\end{equation}

\item  Repeat step 2 and compare the result with the previous iteration:
\begin{equation}
f^{(\eta+1)} \stackrel{?}{\approx} f^{(\eta)}.
\end{equation}
\end{enumerate}

\begin{figure}
[tb]
\centering
\includegraphics[width=0.47\textwidth]{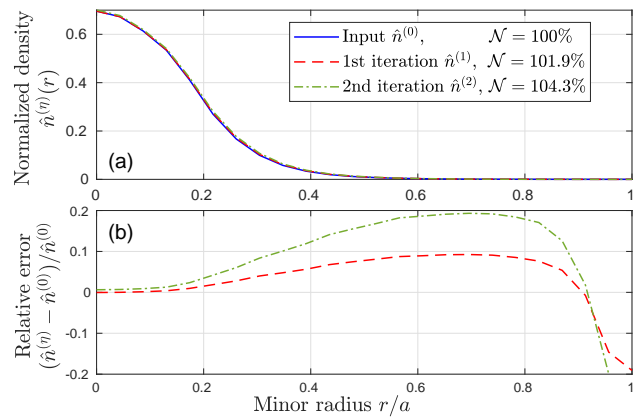}
\caption{(a) Radial density profiles $\hat{n}^{(\eta)}(r)$ with iteration index $\eta = 0,1,2$, and (b) their relative errors in the flat-energy case $E = 0...3.5\,{\rm MeV}$. Unlike in Fig.~\ref{fig:viz_profs_orbits}, the radial profiles in panel (a) here are all normalized by the same value. The total number $\N$ of GCs increases slightly with each iteration as shown in the legend.}
\label{fig:map_0-3500k_dens}%
\end{figure}

\begin{figure}
[tb]
\centering
\includegraphics[width=0.47\textwidth]{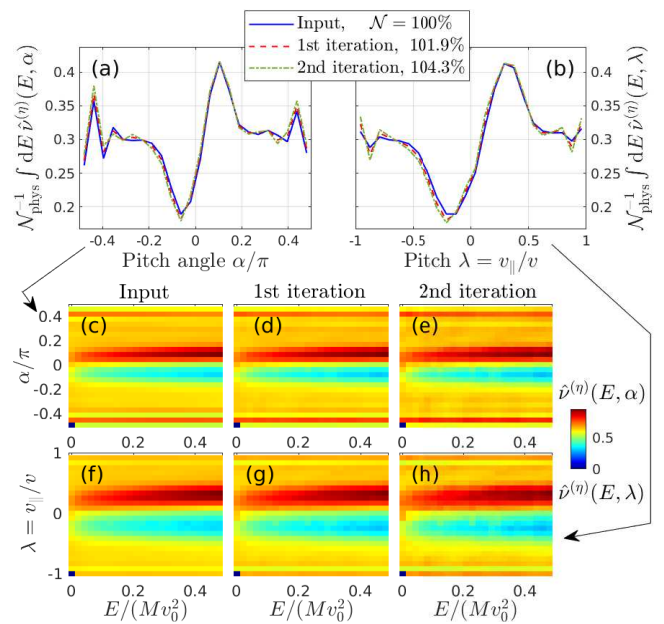}
\caption{Spatially integrated pitch and energy distributions for the cases in Fig.~\protect\ref{fig:map_0-3500k_dens}. The curves in (a) were normalized by the number of GCs $\N$ at each iteration (shown in the legend) in order to highlight the degree of accuracy at which their shape is reproduced. The contours in (c)--(f) are colored differently from Figs.~\protect\ref{fig:viz_0-E0_v-space} and \protect\ref{fig:viz_0-3500k_0th-1st-order} to improve the visibility of small deviations. Usage of 0th-order (binary) binning is responsible for the aliasing noise being larger in panel (a) here than in Fig.~\protect\ref{fig:viz_falpha_mono-E}(a), where a 1st-order (linear interpolation) scheme was used.}
\label{fig:map_0-3500k_pitch}%
\end{figure}

\noindent We have then iterated steps 3 and 4 one more time, looking for increasing deviations that may hint at systematic errors. The results are summarized in Figs.~\ref{fig:map_0-3500k_dens} and \ref{fig:map_0-3500k_pitch}.

Figure~\ref{fig:map_0-3500k_dens}(a) shows that the overall shape of the radial density profile is reproduced well. A closer inspection of the relative errors in Fig.~\ref{fig:map_0-3500k_dens}(b) reveals that the first iteration changes the particle density by about $+2\%$ in the densely populated region $r/a \lesssim 0.3$. Around $r/a \approx 0.7$ the increase is about $+10\%$, and near the boundary we have a reduction by about $-20\%$. The relative error roughly doubles in the second iteration, indicating that it is not random but systematic. As indicated in the legend of Fig.~\ref{fig:map_0-3500k_dens}(a), the total number of GCs $\N$ increases by $1.9\%$ in the first and $2.4\%$ in the second iteration, where the error accumulates to $4.3\%$.

Figure~\ref{fig:map_0-3500k_dens}(b) shows that our procedure has the tendency to increase the particle density in the plasma interior and reduce it near the plasma boundary. We have not investigated why the systematic errors have this particular form and it may, in fact, depend on the particular magnetic configuration at hand. However, the larger errors in the outer region of the plasma are likely related to the fact that, in our test case, this region is populated almost exclusively by particles on large orbits near the trapped-passing boundary (cf.~Fig.~\ref{fig:viz_profs_orbits}(c)), where the poloidal transit time $\tau_{\rm pol}$ exhibits singular behavior (see Fig.~\ref{fig:com_increment_40}) and inaccuracies can be expected. In the present case, the magnitude of these errors can be reduced by increasing the number of phase space samples as shown in Appendix~\ref{apdx:num_converg}.

Figure~\ref{fig:map_0-3500k_pitch} shows the results for the velocity space distributions $\hat{\nu}^{(\eta)}(E,\alpha)$ and $\hat{\nu}^{(\eta)}(E,\lambda)$ for iterations $\eta = 0,1,2$. No significant differences can be seen in the energy direction, which is to be expected since it is a conserved quantity and our energy bins are identical in both representations. But even along the pitch coordinates $\alpha$ and $\lambda$, excellent agreement is obtained after dividing out the systematic increase of $\N$. The noise at large pitch values retains its form as expected for an aliasing effect, and its magnitude increases systematically from one iteration to the next. The minimum associated with the loss cone tends to deepen with each iteration, which is also to be expected due to the irreversibility of losses. As was noted in connection with Fig.~\ref{fig:viz_0-3500k_0th-1st-order}, we see again in Fig.~\ref{fig:map_0-3500k_pitch} that binning on a uniform mesh in $\lambda$ instead of $\alpha$ reduces aliasing and, thus, yields smoother results in the deeply passing domain ($|\alpha| \gtrsim 0.3\pi$ or $|\lambda| \gtrsim 0.8$).

Before proceeding to the conclusion of this paper, we shall add a few encouraging words for readers who may find the results in this section somewhat unsatisfactory. First, it should be noted that we have chosen a fairly challenging setup for our application example: Our density peak is localized deeply in the plasma core, where derivatives of the poloidal magnetic flux $\Psi_{\rm P}$ are small so that $P_\zeta$ is dominated by drift terms. This means the so-called ``non-standard'' orbits are actually prevalent here. Second, we have sampled the CoM space only relatively sparsely (Table~\ref{tab:parm_mesh}) and used no smoothing procedure whatsoever. Even the binning of particles was done using a 0th-order scheme, which is why the noise in Fig.~\ref{fig:map_0-3500k_pitch} is larger than in Fig.~\ref{fig:viz_falpha_mono-E}.

In other words, one may say the we have been looking here at a worst-case scenario with the least degree of sophistication. This should be taken in a positive way, because it means that there is much room for improvement even with well-established techniques. Higher resolution in CoM space, higher-order binning schemes and other smoothing techniques are obvious solutions to reduce noise to a level where the distribution becomes sufficiently smooth to measure local phase space gradients of a particle distribution. The application of such techniques in combination with the orbit-based representation method described here is left for future work.

\section{Summary and conclusion}
\label{sec:discuss}

This paper was motivated by the need to model and process distributions of charged particles in tokamak plasmas in a unified framework like the ITER Integrated Modelling \& Analysis Suite (IMAS), which has to address the needs of a diverse community of users dealing with experimental and numerical data. One of the tasks IMAS will need to perform is the conversion of distribution functions between various coordinates. Our proposal is to construct a database of unperturbed GC drift orbits and use it as a basis for representing distribution functions in constants-of-motion (CoM) space. A working prototype of a concrete workflow as implemented in the code {\tt VisualStart} \cite{BierwageVSTART12,Bierwage12a} is described here.

It should be noted that dedicated IMAS data structures for storing GC orbit properties and quantities along GC trajectories of marker distributions have recently been added into the `distributions' IDS (Interface Data Structures) \cite{ITER_IDS}, making IMAS readily compatible with the representation and methods proposed here.

An orbit-based representation maps any input distribution to an exact equilibrium that is consistent with the specified magnetic field configuration. While the result may generally differ from the original input, the equilibrium constraint ensures that all subsequent mappings from 3-D CoM orbit space back to arbitrary 4-D sets of coordinates become unique and straightforward. In fact, further operations like visualization or exporting are reduced to binning marker particles on user-defined meshes in arbitrary coordinates and arbitrary subspaces.

A new orbit database needs to be set up for each MHD equilibrium, and one may even have to set up one for different energy ranges (say, $1...\text{tens}\,{\rm keV}$, and $0.1...\text{few}\,{\rm MeV}$). The required computational effort should be manageable if the code is optimized and capable of exploiting state-of-the art computing technology. Once the database has been constructed, it can be used to represent and model various populations of charged particles that may be present in the same plasma, such as beam ions, RF-heated ions and fusion products. One orbit database may be shared between particle species with identical charge-to-mass ratio $Qe/M$ by scaling their energies.

The time spent on preparing the orbit database is also justified by the fact that the acquired information can be used for other important tasks, such as resonance analyses and modeling, which must also be part of the IMAS toolbox. The degrees of freedom for modeling are vast, since each orbit can be assigned an independent weight. Rather than prescribing parametric functions of CoM --- whose limitations were briefly discussed in Section~\ref{sec:mdl} --- it is worth exploring unconventional ways that make better use of the degrees of freedom available and take advantage of the fact that our orbit-based representation automatically enforces the equilibrium condition and physical constraints associated with toroidal geometry; namely, the mirror force, magnetic drifts and the resulting velocity space nonuniformities and boundary losses. As an example, we have demonstrated in Section~\ref{sec:viz} how the orbit database can be used to model a distribution of alpha particles with a core-localized birth profile that is not a function of the fast alpha's CoM. This method to design distribution functions was inspired by orbit-following Monte-Carlo simulations, but is much cheaper computationally. We expect that similar strategies can be used to model distribution functions based on various types of experimentally measured data or from results of bounce-averaged Fokker-Planck codes that ignore magnetic drifts \cite{Bierwage17c}.

Finally, we demonstrated in Section~\ref{sec:viz_map} how a distribution of energetic alpha particles with large magnetic drifts can be converted back and forth between a 4-D mesh in arbitrary coordinates and the orbit-based representation in 3-D CoM space. Besides being an essential part of the coordinate transformation toolbox needed by platforms like IMAS, this procedure served us as a method to verify the numerical accuracy and logical integrity of the workflow. In fact, this procedure enabled us to identify and correct subtle problems that arose in early versions of our implementation of the algorithm. The lessons we learned are documented in this paper.

The methods presented here can serve as a starting point that already provides much of the required functionality required by IMAS. One of the next steps should be the adoption or development of smoothing techniques with the goal to mitigate the effect of discretization noise, while minimizing systematic errors associated with such irreversible data manipulations. This is a prerequisite for performing stability analyses of resonant modes and for preparing initial conditions for delta-$f$ simulations, which require explicit information about gradients in the phase space density.

We suggest the reader to also follow related and complementary developments in orbit tomography \cite{Stagner22, StagnerPhdThesis, Jaerleblad21, BenjaminBachelorThesis}. For instance, those studies demonstrate an alternative way to sample the orbit space and employ methods for adaptive meshing and reducing binning errors.

\section*{Acknowledgments}

One of the authors (A.B.) acknowledges insightful discussions with Yasushi Todo concerning the handling of the Jacobian factor $B^*_\parallel$, and with Stuart Benjamin and David Pfefferle concerning methods for phase space discretization and mapping techniques. On a broader note, A.B.\ would like to thank Liu Chen, Zhihong Lin, Fulvio Zonca as well as Sergio Briguglio, Claudio DiTroia and Gregorio Vlad for their support during the early development stages of {\tt VisualStart} in 2008--2010 for the hybrid code {\tt HMGC}. Subsequent extensions enabling {\tt VisualStart} to initialize long-time simulations performed for JT-60U with the code {\tt MEGA} benefited from support by Kouji Shinohara, Yasushi Todo and Masatoshi Yagi, to whom A.B.\ would also like to express his gratitude. Finally, A.B.\ is grateful to Jeronimo Garcia and Shunsuke Ide for encouraging and supporting a fruitful collaboration between JET and QST. The working example used in the present paper was inspired by plasmas produced in recent JET experiments with central 3-ion RF heating, which posed a suitably challenging and practically relevant scenario for testing our numerical algorithms and workflow. We would like to thank all JET contributors for making these interesting experiments possible and allowing us to participate.

The workstation used for the numerical calculations reported here was funded by QST President's Strategic Grant (Creative Research). This work has been partially carried out within the framework of the EUROfusion Consortium and has received funding from the Euratom research and training programme 2014-2018 and 2019-2020 under Grant Agreement No.\ 633053. The views and opinions expressed herein do not necessarily reflect those of the European Commission.

\addcontentsline{toc}{section}{Appendices with supplementary information}
\appendix
\addtocontents{toc}{\setcounter{tocdepth}{-1}}

\section{Binning operations and Jacobians}
\label{apdx:bin}

In this section, we describe how we evaluated the Klimontovich representation (\ref{eq:f_w}) to visualize our results in Section~\ref{sec:viz}. For clarity, we break the procedure up into (i) the computation of a histogram $h$ and (ii) its conversion to a density function $f$. As a concrete example, we use the set of noncanonical coordinates $(E,\lambda,R,z)$. The cell center positions are ${\bm Z}_{ijkm} = (E_i,\lambda_j,R_k,z_m)$ and the volume elements are $\Delta V_{ijkm} = \Delta E_i\Delta\lambda_j\Delta R_k\Delta z_m$.\footnote{In contrast to the canonical 6-D phase space volume elements $\Delta\V_{\rm orb}$ and $\Delta\V$ of GC orbits and their markers as defined in Section~\ref{sec:weight}, the symbol $\Delta V$ appearing here denotes an element in an arbitrary mesh in an arbitrary number of dimensions.}
The corresponding histogram of GCs is given by
\begin{equation}
h({\bm Z}_{ijkm}) = \sum_{l=1}^{N_{\rm mrk}} w_l \frac{S({\bm Z}_{{\rm gc},l} - {\bm Z}_{ijkm})}{\Delta V_{ijkm}}.
\label{eq:h_map}
\end{equation}

\noindent Marker $l$ is located at GC position ${\bm Z}_{{\rm gc},l}$, and its weight $w_l$ determines how many physical particles this marker is meant to represent (cf.~Eq.~(\ref{eq:w_f})). The shape function $0 \leq S \leq 1$ prescribes how we map these weights onto our mesh. The phase space density function $f$ is then
\begin{equation}
f = \frac{h}{\J^{({\bm x},{\bm v})}_{\rm gc}} \approx \frac{h({\bm Z}_{ijkm})}{\left[\frac{B^*_\parallel}{B} \J^{\bm x}_{Rz} \J^{\bm v}_{E\lambda}\right]_{ijkm}};
\label{eq:f_h_map}
\end{equation}

\noindent where $\J^{({\bm x},{\bm v})}_{\rm gc}$ is the Jacobian for the conversion from Cartesian to noncanonical GC coordinates. Substituting Eq.~(\ref{eq:h_map}) into (\ref{eq:f_h_map}) and assuming, within the accuracy limits of our discrete mesh, that $[...]_{ijkm} \approx [...]_l$, the formula for mapping the phase space density function $f$ to our 4-D mesh becomes\footnote{Conventional PIC codes sample the phase space by loading marker particles in a uniformly randomized manner at the beginning of a simulation ($t = 0$). In that case, the initial marker weights (via their volume elements) carry a factor $B^*_\parallel(t=0)/B(t=0)$. When evaluating moments of the distribution function during a simulation, that factor is effectively multiplied by the inverse factor $B(t)/B^*_\parallel(t)$ that appears in Eq.~(\protect\ref{eq:f_map}). These two factors cancel approximately and may, thus, be omitted, if one ignores corrections of order $\O(\rho_0^2/a^2)$.}
\begin{equation}
f(E_i,\lambda_j,R_k,z_m) = \sum_{l=1}^{N_{\rm mrk}} \frac{w_l \times S({\bm Z}_{{\rm gc},l} - {\bm Z}_{ijkm})}{ \left[\frac{B^*_\parallel}{B} \J^{\bm x}_{Rz} \J^{\bm v}_{E\lambda}\right]_l \Delta V_{ijkm}}.
\label{eq:f_map}
\end{equation}

Based on Eq.~(\ref{eq:f_map}), we compute density fields in 1-D and 2-D by binning marker weights $w_l$ as follows:
\begin{align}
n(\psi_{{\rm P},i}) =& \sum_{l=1}^{N_{\rm mrk}} \frac{w_l}{\left[\frac{B^*_\parallel}{B} \J^{\bm x}_{\psi_{\rm P}}(R,z)\right]_l} \frac{S({\bm Z}_{{\rm gc},l} - {\bm Z}_i)}{\Delta \psi_{{\rm P},i}},
\label{eq:f_r}
\\
n(R_i,z_j) =& \sum_{l=1}^{N_{\rm mrk}} \frac{w_l}{\left[\frac{B^*_\parallel}{B} \J^{\bm x}_{Rz}(R)\right]_l} \frac{S({\bm Z}_{{\rm gc},l} - {\bm Z}_{ij})}{\Delta R_i\Delta z_j}.
\label{eq:f_rz}
\end{align}

\noindent The spatially integrated velocity distributions are
\begin{align}
\nu(a_i,b_j) =& \sum_{l=1}^{N_{\rm mrk}} \frac{w_l}{\left[\frac{B^*_\parallel}{B}\hat{\J}_{ab}^{\bm v}\right]_l} \frac{S({\bm Z}_{{\rm gc},l} - {\bm Z}_{ij})}{\Delta a_i \Delta b_j}.
\label{eq:f_v}
\end{align}

\noindent In Sections~\ref{sec:viz_monoE} and \ref{sec:viz_flatE}, we used 1st-order shape functions $S$ for binning (linear interpolation), and in Section~\ref{sec:viz_map} we used 0th-order (binary) binning.

Finally, we present expressions for the Jacobian factors multiplying $B^*_\parallel/B$. Converting the right-hand side of Eq.~(\ref{eq:dvdx_noncan}) to polar GC velocity coordinates by substituting ${\rm d}\mu \rightarrow B^{-1}M{\rm d}v_\perp^2/2$, we obtain
\begin{equation}
{\rm d}\hat{v}_x{\rm d}\hat{v}_y{\rm d}\hat{v}_z\times{\rm d}^3{\bm x} = \frac{B^*_\parallel}{B}\times 2\pi \hat{v}_\perp {\rm d}\hat{v}_\perp {\rm d}\hat{v}_\parallel \times {\rm d}^3{\bm x}.
\end{equation}

\noindent The Jacobian $\J^{\bm x}_{\psi_{\rm P}}$ for transforming between ${\rm d}^3{\bm x}$ and the normalized poloidal flux $\psi_{\rm P}$ is evaluated numerically. The Jacobian for cylinder coordinates is $\J_{Rz}^{\bm x} = 2\pi R$. Analytical expressions can also be derived for the Jacobians of all velocity coordinates that we use here. Namely, from the relations
\begin{align}
\hat{v}_\perp^2 =& 2 \hat{E} \Lambda \hat{B} = 2 \hat{E} \cos^2\alpha = 2 \hat{E} (1 - \lambda^2), \\
\hat{v}_\parallel =& \sqrt{2\hat{E}}\sin\alpha = \sqrt{2\hat{E}}\lambda = \sigma_B\sqrt{2\hat{E}(1 - \Lambda\hat{B})},
\end{align}

\noindent with $\sigma_B \equiv v_\parallel/|v_\parallel|$, one readily obtains the Jacobians
\begin{equation}
\J_{ab}^{\bm v} = 2\pi \left|
\begin{array}{cc}
\frac{1}{2}\partial_a v_\perp^2 & \partial_a v_\parallel \\
\frac{1}{2}\partial_b v_\perp^2 & \partial_b v_\parallel
\end{array}\right|
\end{equation}

\noindent for locally (in ${\bm x}$) transforming velocity space elements $2\pi v_\perp{\rm d}v_\perp{\rm d}v_\parallel$ to other sets $(a,b)$, such as
\begin{gather}
\hat{\J}_{\enr\mu}^{\bm v} = \pi\hat{B}/|\hat{v}_\parallel|, \quad
\hat{\J}_{\enr\Lambda}^{\bm v} = 2\pi{\hat{E}\hat{B}}/|\hat{v}_\parallel|,
\label{eq:jv1}
\\
\hat{\J}_{E \alpha}^{\bm v} = \hat{\J}_{v_\perp,v_\parallel} = 2\pi \hat{v}_\perp, \quad
\hat{\J}_{E \lambda}^{\bm v} = 2\pi\hat{v}.
\label{eq:jv2}
\end{gather}

\section{Numerical accuracy and data storage}
\label{apdx:num}

Researchers tend to invest significant effort in the preparation of their simulation scenarios, balancing accuracy against speed. However, the data on platforms like IMAS are likely to be of variable quality, since some of it originates from automated or semi-automated workflows. The tools used to process those data should hence have a certain degree of tolerance with respect to inaccuracies and be able to detect and handle exceptions. The workflow in {\tt VisualStart} also contains such sensibility checks and ways to mitigate the effects of inaccuracies. Some of the inaccuracies arising in {\tt VisualStart} could be avoided with more sophisticated techniques, while others are practically inevitable due to numerical resolution being necessarily finite. This Appendix contains notes concerning some issues that are relevant for the subject of this paper or, at least, for the current implementation of {\tt VisualStart}.

\subsection{Particle pushing}

At present, the particle pushing routine used to compute GC orbits in {\tt VisualStart} uses the same finite-difference algorithm as the MHD-PIC hybrid code {\tt MEGA} \cite{Todo98,Todo05}: a 4th-order Runge-Kutta scheme advances the equations of motion expressed in cylinder coordinates. Spatial derivatives are also taken through simple finite differences. This choice has been made deliberately in order to have the same level of accuracy as the hybrid code for which {\tt VisualStart} prepares initial conditions. This allows us to detect possible problems, if any, at an early stage, before launching expensive hybrid simulations. Of course, accuracy can be improved with various techniques, including a Hamiltonian formulation like that used in codes like {\tt ORBIT} \cite{White84,ChenY99,White19} and {\tt HAGIS} \cite{Pinches98, Pinches04}, albeit with the complication of having to work with a specialized set of coordinates. The benefits of such techniques become noticeable when following orbits for long periods of time (as in Poincar\'{e} analyses). For evaluating a single poloidal transit as we do here, simple techniques often suffice, but due attention must always be paid to regions near singularities and stagnation points like those shown in Figs.~\ref{fig:tp-boundary} and \ref{fig:com_increment_40}.

\begin{figure}
[tbp]
\centering
\includegraphics[width=0.47\textwidth]{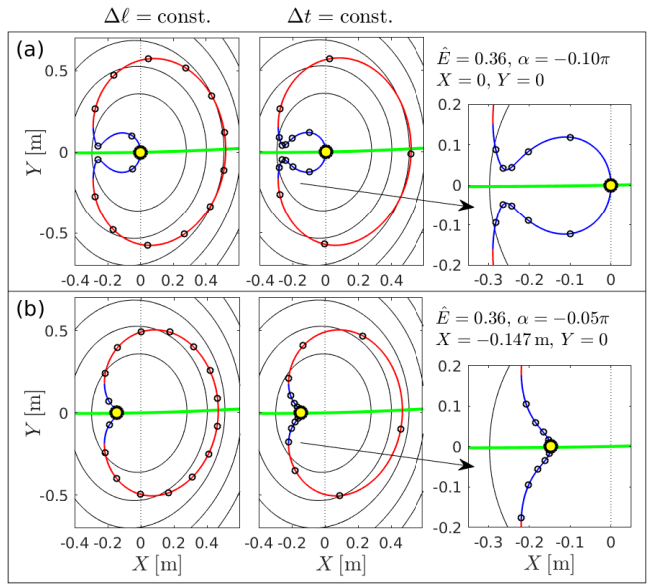}
\caption{Sampling of potato orbits. The two examples in (a) and (b) correspond to $2.52\,{\rm MeV}$ alpha particles with different pitch angles in the scenario of Fig.~\protect\ref{fig:tok} that is partly based on JET. The coordinate values listed are those at the starting point of an orbit contour, which is indicated by a large yellow circle. The small black circles represent $N_\tau = 16$ samples separated by equal length intervals ($\Delta\ell = L_{\rm orb}/N_\tau$, left) or equal time intervals ($\Delta t = \tau_{\rm orb}/N_\tau$, central and right column).}
\label{fig:sample_dtdl}%
\end{figure}

\subsection{Orbit database}

Although one may load GC phase space markers as soon as an orbit contour has been computed, it is advantageous to store orbits in a database and load the markers later as needed. One reason is that the orbits can be recycled when preparing distribution functions for different sets of particles: alpha particles and deuterons are indistinguishable for identical velocities, and one may have multiple beams as well as RF-heated populations of a certain particle species in the same plasma. These sets of particles may all share the same orbit database.

In order to keep the size of the database manageable, the representation of the orbits must be efficient. {\tt VisualStart} saves all orbits in discretized form, typically using 32--64 samples per orbit. The user can choose between three options for the distribution of samples: (i) uniform in space, (ii) uniform in time, or (iii) a 50:50 combination of (i) and (ii). We usually choose option (i) for beam-like distributions that consist only of passing particles. Option (iii) is preferred when trapped particles are present in order to ensure good resolution both near and far away from V-type stagnation points as the examples in Fig.~\ref{fig:sample_dtdl} show.

The recorded time arrays must include at least $R(t)$, $Z(t)$, $v_\parallel(t)$ and the time $t$ itself. Our databases contain many other arrays that have been stored for various purposes, such as $\zeta(t)$, $v_R(t)$, $v_z(t)$, $v_\zeta(t)$ and $\psi_{\rm P}(t)$. The database file from our low-resolution example in Fig.~\ref{fig:mesh} has a size of 33 Mbyte in {\tt NetCDF} format. The databases with the parameters in Table~\ref{tab:parm_mesh} occupy about 150 Mbyte each in the mono-energetic cases and 500 Mbyte in the flat-energy cases. These files also contain useful auxiliary information, such as the numerical parameters, poloidal and toroidal transit times, the orbit length, bounce angle, the orbit type, and some measures of the computational accuracy: energy conservation, and spatial mismatch between start- and end-point.

Orbits that have not been completed within a certain upper bound of time steps are labeled as ``incomplete''. Before proceeding to Step 4 of our workflow in Fig.~\ref{fig:vstart}, we deal with such incomplete orbits and those that have insufficient accuracy. Bad orbits can be corrected if necessary, or simply discarded if sufficient accuracy is difficult to achieve and unnecessary (as may happen for tiny orbits very close to an O-type stagnation point).

Data related to marker weighting are also present; in particular, the volume element sizes $\Delta\V_{\rm orb}$. There is an option to undo the factor $1/2$ in the volume element given by Eq.~(\ref{eq:vorb}) for stagnation orbits that are smaller than the cell size and, hence, cannot be double-counted. Their volume elements may then be rescaled in proportion to the orbit's radial diameter, but the effect is usually negligible since such orbits carry little weight (and one should use a finer mesh if they are of interest).

\begin{figure}
[tbp]
\centering
\includegraphics[width=0.45\textwidth]{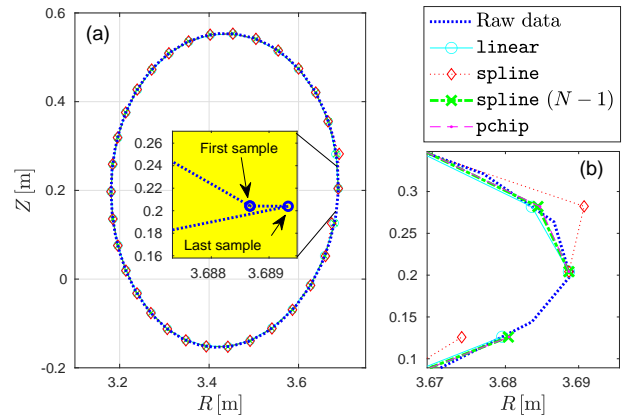}
\caption{Example of interpolation errors caused by a small mismatch between the start and end points of an orbit, which can be seen in the zoomed inset in panel (a). Panel (b) shows the results of various interpolation algorithms using the {\tt Matlab} function {\tt interp1}: {\tt linear} interpolation, piecewise cubic spline ({\tt spline}), {\tt spline} without the last sample (``$N-1$''), and shape-preserving piecewise cubic interpolation ({\tt pchip}). The magnetic field in this example is based on a JT-60U plasma as used, for instance, in Ref.~\protect\cite{Bierwage18}.}
\label{fig:interpol}%
\end{figure}

Interpolation is necessary when converting the raw data from the particle pushing algorithm to the samples used to represent an orbit in the database. Further interpolation is necessary when loading markers on an orbit that has been retrieved from the database. As shown in Fig.~\ref{fig:interpol}, we found that a tiny discontinuity in an orbit contour (mismatch between start and end points) can cause relatively large ``ringing'' in spline interpolations. Subsequent interpolations in that region --- namely, when we load marker particles --- can produce corrupt samples that may even lie outside the plasma. When applying the 1-D interpolation function ${\tt interp1}$ of {\tt Matlab} for processing our orbits (and other potentially ``noisy'' data), we avoid such problems by using the {\tt pchip} option instead of {\tt spline}.

\subsection{Marker loading}

In order to avoid crowding at the midplane, the markers on each orbit are loaded with a random time offset $0 \leq \tau_0 < \tau_{\rm pol}$ relative to the starting point at the midplane. When using the markers in a simulation, they are also spread (randomly) along the toroidal angle $\zeta$.

\begin{figure}
[tbp]
\centering
\includegraphics[width=0.45\textwidth]{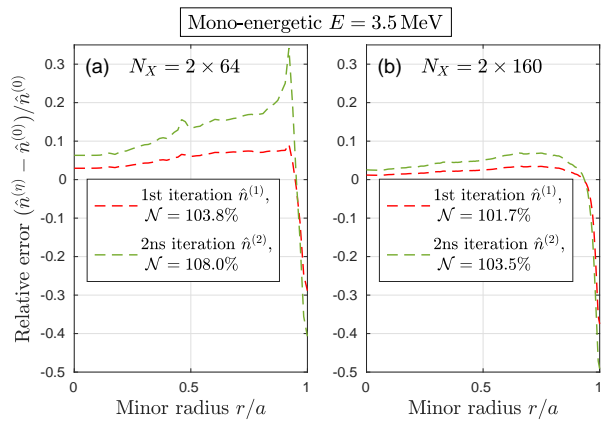}
\caption{Convergence test showing the reduction of the relative errors in the radial density profiles $\hat{n}^{(\eta)}(r)$ with iteration index $\eta = 1,2$. The analysis is equivalent to that in Fig.~\protect\ref{fig:map_0-3500k_dens}, except that it is performed here for the monotonic-energetic case with $E = 3.5\,{\rm MeV}$. Panel (a) shows the result for the default number of $N_X = 2\times 64$ radial grid points. In panel (b) it is $N_X = 2\times 160$. The other parameters are the same as in Table~\protect\ref{tab:parm_mesh}, except that the number of markers per orbit $N_\tau$ was increased from $48\tfrac{L_{\rm orb}}{L_{\rm bnd}}$ to the same value of $480\tfrac{L_{\rm orb}}{L_{\rm bnd}}$ as used for Fig.~\protect\ref{fig:map_0-3500k_dens}. The relative change in the total number $\N$ of GCs is shown in the legends.}
\label{fig:converg}%
\end{figure}

\subsection{Convergence test}
\label{apdx:num_converg}

Figure~\ref{fig:map_0-3500k_dens}(b) showed indications of relatively large errors that increased systematically with each iteration of our back-and-fourth transformation between a 4-D mesh in the coordinates $(E,\lambda,R,z)$ and the orbit-based representation in 3-D CoM space. In our discussion in Section~\ref{sec:viz_map}, we attributed this problem to inaccuracies in the estimation of the volume elements for orbits sampling the region near the trapped-passing boundary, where some factors exhibit singular behavior.

Further tests show that these errors can be reduced by simply increasing the number of phase space samples via the parameters $N_\alpha$, $N_X$ and $N_\tau$ in Table~\ref{tab:parm_mesh}.\footnote{Since kinetic energy is conserved and our 3-D and 4-D distributions both share the same energy mesh, the parameter $N_E$ has no impact on the numerical accuracy in our verification exercise.}
In Fig.~\ref{fig:converg} we demonstrate this for the case where we increased the number of radial grid points $N_X$. While the flat-energy case with $E = 0...3.5\,{\rm MeV}$ was used in Fig.~\ref{fig:map_0-3500k_dens}, we have chosen only mono-energetic alpha particles with $E = 3.5\,{\rm MeV}$ for the convergence test in Fig.~\ref{fig:converg}. For an identical mesh, this results in larger values of the relative errors in Fig.~\ref{fig:map_0-3500k_dens}, presumably because the signal-to-noise ratio is reduced.

The comparison between panels (a) and (b) of Fig.~\ref{fig:converg} shows a significant reduction of the numerical errors in both the local values of the alpha particle density profile $\hat{n}^{(\eta)}(r)$ and the total number of GCs $\N$. As before, the errors accumulate with each iteration, which confirms their systematic character. Large errors persist only at the artificial loss boundary ($r/a \approx 1$), which is to be expected and requires dedicated solutions that are not covered here.

Besides increasing the number of samples, there may exist other ways to correct or otherwise suppress the inaccuracies associated with singular volume elements. This is left for future work. A preliminary discussion was given in Section~\ref{sec:weight_discuss} of the main text.


\bibliographystyle{unsrt}
\bibliography{references}

\end{document}